\documentstyle [prb,aps,epsf]{revtex}
\begin{document}
\widetext
\draft

\title{
Quantum Monte Carlo study
of the one-dimensional Holstein model of
spinless fermions
}

\author{Ross H. McKenzie\cite{email}, C. J. Hamer,
 and D. W. Murray}                        

\address{School of Physics, University of New
South Wales, Sydney 2052, Australia}

\date{Received 29 November 1995}
\maketitle
\begin{abstract}
The Holstein model of spinless fermions interacting with
dispersionless phonons in one dimension is studied by a Green's
function Monte Carlo technique.  The ground state energy,
first fermionic excited state, density wave correlations,
and mean lattice displacement are calculated for lattices
of up to 16 sites, for one fermion per two sites, i.e., a
half-filled band. Results are obtained for values of the fermion
hopping parameter of $t=0.1 \omega$, $\omega$,  and 
$10 \omega$    where  $\omega$  is the
phonon frequency.  At  a finite fermion-phonon coupling $g$ 
there is a transition from a metallic phase to an insulating
phase in which there is charge-density-wave order.  Finite size
scaling is found to hold in the metallic phase and is used to
extract the coupling dependence of the Luttinger liquid
parameters, $u_\rho$ and $K_\rho$, the velocity of charge excitations
and the correlation exponent, respectively.  For              
free fermions ($g=0$) and for strong coupling ($g^2 \gg t \omega$)
our results agree well with known analytic results.
For $t=\omega$ and $t=10\omega$ our results are inconsistent
with the metal-insulator transition
being a Kosterlitz-Thouless transition.\\
\\
To appear in Physical Review B, April 15, 1996.\\
\end{abstract}



\section{Introduction}
\label{secintro}

A wide range of quasi-one-dimensional materials have electronic
properties that are dominated by the Peierls or
charge-density-wave instability caused
by the electron-phonon interaction \cite{gru}.
For a
half-filled band it is energetically favourable for the lattice
to dimerize and open an energy gap at the Fermi surface. Although
the lattice distortion increases the lattice
energy, opening the electronic energy gap preferentially lowers the total
energy for highly anisotropic systems \cite{pei}.  These systems are often
modelled by the one-dimensional Holstein \cite{hols}
 or Su-Schrieffer-Heeger (SSH) \cite{hee}
models.  Most treatments of the Peierls
instability treat the phonons in
the mean-field or rigid lattice approximation.  This is questionable
in one-dimension and furthermore, in a wide-range of materials
the lattice distortion is comparable to the zero-point motion of
the lattice \cite{mck}.  It has recently been shown that the quantum lattice
fluctuations must be taken into account to satisfactorily describe
optical properties \cite{kim,lon}. 
Several authors have previously considered the 
role of quantum lattice fluctuations
for  the SSH model \cite{fra}
and the Holstein model \cite{hir,bou,zhe,wu} at half-filling.
Voit and Schulz have considered the interplay of
quantum lattice fluctuations and electron-electron
interactions away from half-filling \cite{voit2}.
Recently the Holstein model, also known as the molecular crystal model,
has received considerable attention because
the challenge of high-$T_c$ and fullerene
superconductors has revealed
deficiencies in our understanding of the electron-phonon interaction
and the competition between superconductivity and
charge-density-wave instabilities.
This has motivated studies of the Holstein model
in infinite dimensions \cite{fre}, two dimensions \cite{gub},
one dimension \cite{mar}
and on just a few sites \cite{ale}.

We consider the Holstein model in one dimension at half filling
and only with spinless fermions, for simplicity. 
The spinless fermions hop 
along a one-dimensional chain and
interact with a phonon mode located
on each lattice site.
The creation operator for a fermion                        
on site $i$ is denoted $ c_i$.
The fermions can hop between neighbouring sites with
amplitude $t$.
In the absence of interactions
the phonons all have the same frequency $\omega$,
i.e., they are dispersionless.
The electron-phonon coupling, in units of energy, is $g$.
Phonon position and momentum    
operators are denoted by $q_i$ and $p_i$, respectively.
The Hamiltonian for the Holstein model (at half filling) is \cite{hir}
\begin{equation}
H= -t \sum_i \left(c_i^\dagger c_{i+1} + c_{i+1}^\dagger c_i \right)
   - g (2M \omega)^{1/2} \sum_i (c_i^\dagger c_i - {1 \over 2}) q_i
 + \sum_i {1 \over 2 M} p_i^2 + {1 \over 2} M \omega^2 q_i^2
-{N \over 2} \omega
   \label{ab1}
\end{equation}
for a system of $N$ lattice sites.
This Hamiltonian has particle-hole symmetry since
the transformation $c_i \to (-1)^i c_i^\dagger$,
$q_i \to -q_i$ leaves $H$ invariant.
This discrete symmetry is broken in the
charge-density-wave phase which has the electronic order parameter
\begin{equation}
m_e \equiv { 1 \over N} \sum_i (-1)^i <c_i^\dagger c_i>
   \label{op1}
\end{equation}
and the phonon order parameter
\begin{equation}
m_p \equiv { 1 \over N} \sum_i (-1)^i <q_i>
   \label{op2}
\end{equation}
which is a measure of the dimerization.

If phonon creation and annihilation
operators are denoted by $a_i^\dagger$ and $a_i$, respectively,
the Hamiltonian (\ref{ab1}) can be written
\begin{equation}
H= -t \sum_i \left(c_i^\dagger c_{i+1} + c_{i+1}^\dagger c_i \right)
   - g \sum_i (c_i^\dagger c_i - {1 \over 2})
        \left(a_i + a_i^\dagger \right)
   + \omega \sum_i a_i^\dagger a_i.
   \label{abx1}
\end{equation}
Thus ground state properties will be determined by two
independent parameters, which we will take to be
$t/\omega$ and $g/\omega$.
It is also useful to define the dimensionless
electron-phonon coupling 
\begin{equation}
\lambda \equiv {g^2 \over \pi t \omega}.
   \label{ac1}
\end{equation}

Although for simplicity we confine ourselves to the case of
spinless fermions this model is still of physical relevance
in at least two situations.
The first situation concerns 
strongly correlated electron systems.
In the infinite $U$ limit the Hubbard model is known to map
onto the case of spinless fermions \cite{voit}.
This may be realized in the 1:2 TCNQ salts \cite{ric}.
The second situation concerns the spin-Peierls transition \cite{sp}.
Using a 
Jordan-Wigner transformation \cite{frad} this model can be mapped
onto a $XX$ spin chain in zero field with the
Hamiltonian:
\begin{equation}
H= -2 t \sum_i \left(S_i^x S_{i+1}^x + S_i^y S_{i+1}^y \right)
   - g \sum_i S_i^z \left(a_i + a_i^\dagger \right)
   + \omega \sum_i a_i^\dagger a_i.
   \label{ab2}
\end{equation}
It should be pointed out that this is not the standard 
Hamiltonian used to study the spin-Peierls transition.
However, it does have the same qualitative features:
i.e., a dimerization of the phonons results in a
spin singlet ground state with an energy gap.

It was recently shown \cite{ben} rigorously that for
the one-dimensional Holstein model of spinless fermions
 at half-filling there is
no long range order for sufficiently small coupling $g$.
Hirsch and Fradkin \cite{hir} studied
the Holstein model at half-filling using
a world-line Monte Carlo technique and a strong coupling expansion.
The expansion suggested that for spinless fermions quantum
lattice fluctuations destroy the dimerized state if the
fermion-phonon coupling was sufficiently weak and the phonon frequency
sufficiently high.  The quantum Monte Carlo simulations were performed
for $0.5 \omega < t < 3 \omega$ 
and gave a phase diagram qualitatively consistent
with the strong coupling expansion.  In contrast, for fermions with
spin their results were consistent with dimerization for 
finite phonon frequency and all non-zero couplings.

Zheng, Feinberg, and Avignon \cite{zhe}
 used a variational polaron wave function
to study the Holstein model at half-filling. For spinless fermions
the ground state is a
charge-density wave for {\it all} parameter values.
Most of their results were consistent with
Hirsch and Fradkin. However, they
found that for large phonon frequencies ($t > 0.3\omega$)
there was
a first-order phase transition, with a
very large jump in the CDW order parameter, between CDW phases
 when $g^2 \sim 10 \omega$.
They point out that this transition may be an artefact of the
variational
treatment since it is known that in small-polaron theory 
of a single electron a similar
two-minimum structure, leading to non-analytic behaviour sometimes
referred to as ``self-trapping,'' occurs and is known to be an artefact of the
variational teatment \cite{sho,ger}.

This paper presents a study of the Holstein model
using a Green's Function Monte Carlo technique.
Section \ref{seclutfss} reviews how the
metallic phase should be a Luttinger liquid
and how finite-size scaling can be used to
extract the Luttinger liquid parameters.
Section \ref{secana} briefly summarizes known
analytic results of the Holstein model that
can be used to check and help understand
our Monte Carlo results.
Section \ref{secgfqmc} contains a detailed
description of the 
Green's Function Monte Carlo technique that we use.
Our results are presented and interpreted in
Section \ref{secres}. The physical picture 
that emerges from our results is
discussed in the final section.

\section{Luttinger liquids and Finite-size scaling}
\label{seclutfss}

\subsection{The Luttinger liquid conjecture}
\label{seccon}

For weak coupling and high frequency the system is
in a metallic, i.e., gapless phase.
According to Haldane's ``Luttinger liquid'' conjecture
\cite{voit,hald} this phase should be in the
same universality class as the Tomonaga-Luttinger
model of interacting spinless fermions.
This means the low-energy properties
of the metallic phase are completely described
by an effective Luttinger model with two parameters
$u_\rho$,  the velocity of charge excitations
or renormalized Fermi velocity,
and $K_\rho$, the renormalized effective coupling (stiffness) constant.
Important properties  of the Luttinger model,
quite distinct from those of a conventional
Fermi liquid, are
(i) there are no quasi-particle excitations
at the Fermi surface and (ii)
all correlation functions have non-universal
exponents that can be written in terms of
the single parameter $K_\rho$.
For example,
$K_\rho$ determines the singularity of the
momentum distribution function close to the Fermi surface \cite{voit3}:
\begin{equation}
n(k) \simeq {1 \over 2} - {\rm sign}(k-k_F)|k-k_F|^\alpha
\label{nk}
\end{equation}
and of the single-particle density of states
\begin{equation}
\rho(E) \sim |E|^\alpha
\label{re}
\end{equation}
where
\begin{equation}
\alpha \equiv {1 \over 2}(K_\rho + {1 \over K_\rho} -2).
\label{alpha}
\end{equation}
For free fermions $K_\rho=1$ and $\alpha= 0$.
For attractive (repulsive) interactions $K_\rho>1$ ($K_\rho<1$).
It is remarkable 
that, as explained below, the parameters $u_\rho$ and
$K_\rho$ can be determined from
numerical calculations on systems of finite size.

\subsection{Finite size scaling}
\label{secfin}

The ground state energy $E_0(N)$ of a conformally invariant
 system of $N$ sites is,  to leading order in $1/N$ \cite{aff},
\begin{equation}
{E_0(N) \over N}= \epsilon_\infty - {\pi u_\rho C \over 6 N^2}
\label{e0}
\end{equation}
where $\epsilon_\infty$ is the ground state energy
per site of the
infinite system, $u_\rho$ is the velocity of charge excitations,
and $C$ is the conformal charge.
Care must be taken with boundary conditions.
We use 
anti-periodic (periodic) boundary conditions
for the fermions when there is an even (odd) number
of fermions. This corresponds to periodic 
boundary conditions for the associated spin
or bosonic models \cite{frad}.
If the system is a Luttinger liquid it belongs to the
same universality class as the Gaussian model and $C=1$ \cite{voit}.
The slope of a plot of $E_0(N)/N$ versus $1/N^2$
(compare Figure \ref{fige0}) can then be used to determine
$u_\rho$.

The energy of the first excited state is,
to leading order in $1/N$,
\begin{equation}
E_1(N)-E_0(N) =  {2 \pi u_\rho x \over  N}
\label{e1}
\end{equation}
where $x$ is the scaling dimension \cite{cardy}. 
A Luttinger liquid has the unusual property that 
$x$ depends on the coupling constants.

In the presence of particle-hole symmetry
$x$ can be related to the correlation exponent $K_\rho$
which determines the asymptotic decay 
of {\it all} correlation functions.
Let $E_{\pm 1 }(N)$ denote the ground state energy
of $N/2 \pm 1$ fermions on $N$ sites.
By particle-hole symmetry
 $E_{+1}(N)=  E_{-1}(N).$ 
In a general Luttinger liquid of spinless fermions\cite{voit}
with charge density $n$ 
the compressibility $\kappa$ is given by
\begin{equation}
 { 1 \over n^2 \kappa}
\equiv { \partial^2 \epsilon_\infty(n) \over \partial n^2}
= {\pi u_\rho \over K_\rho} 
\label{lute1}
\end{equation}
Since particle-hole symmetry implies $
 { \partial \epsilon_\infty(n) \over \partial n}= 0$
it follows that
\begin{equation}
E_{\pm 1 }(N)
=E_0(N) + {1 \over 2} \left( {1 \over N}\right)^2 N
 { \partial^2 \epsilon_\infty(n) \over \partial n^2}
\label{lute2}
\end{equation}
which with (\ref{lute1}) implies that to leading order in $1/N$
\begin{equation}
E_{\pm 1 }(N)
-E_0(N) =  {\pi u_\rho  \over 2 K_\rho  N}
\label{es1}
\end{equation}
This is identical to (\ref{e1}) with
 $K_\rho = 1 / 4x$.
Hence, if $u_\rho$ is known a plot of the 
energy gap versus $1/N$ (compare Figure \ref{fige1})
can be used to determine $K_\rho$.

\section{Analytic results}
\label{secana}

Certain limits of the Holstein model for
which analytic results can be obtained are now
briefly reviewed. These results will
be compared to the appropriate numerical results.

\subsection{Localized fermions ($t=0$)}
\label{secloc}

The fermions cannot move between sites and the
Hamiltonian reduces to $N$ independent Hamiltonians.
The Hamiltonian  for the $i$th site is
\begin{equation}
H_i= {1 \over 2 M} p_i^2 + {1 \over 2} M \omega^2 q_i^2
  - (n_i - {1 \over 2}) g (2M\omega)^{1/2} q_i
 -{1 \over 2} \omega
   \label{qb1}
\end{equation}
where $n_i\equiv c_i^\dagger c_i$ is the fermion
occupation at site $i$.
The presence or absence of a fermion shifts the
equilibrium position of the oscillator to
$+q_e$ or $-q_e$, respectively, where
\begin{equation}
q_e=g \left( { M \over 2 \omega} \right)^{1 \over 2}
=<q_i(2n_i-1)>
   \label{qo1}
\end{equation}
This Hamiltonian can be diagonalized by the
Lang-Firsov transformation \cite{mah}:
$c_i \to c_i \exp(q_e(a_i^\dagger - a_i))$,
$a_i \to a_i - (2n_i - 1)q_e$.
The mean square lattice displacement is
\begin{equation}
<q_i^2>=q_e^2+ {1 \over 2 M \omega}.
   \label{q21}
\end{equation}
The ground state energy per site is
\begin{equation}
\epsilon_\infty = - { g^2 \over 4\omega} .
   \label{eio1}
\end{equation}

\subsection{Small Polarons ($g^2 \gg t \omega$)}
\label{secpol}

This corresponds to the case of a narrow band of small
polarons \cite{hols}.
The intersite hopping represents a small 
perturbation on the situation considered in
the previous section.
Hirsch and Fradkin \cite{hir} derived an effective
Hamiltonian \cite{beni} involving the new fermion operator
$C_i^\dagger$ that creates a fermion at site $i$
and changes the oscillator ground state
from one centered at $-q_e$ to one centered at $+q_e$.

\begin{equation}
H_{eff}= -N {g^2 \over 4 \omega}
-J \sum_i \left(C_i^\dagger C_{i+1} + C_{i+1}^\dagger C_i \right)
   + V \sum_i \left(C_i^\dagger C_i - {1 \over 2} \right)
 \left(C_{i+1}^\dagger C_{i+1} - {1 \over 2} \right) 
   \label{effa1}
\end{equation}
The first term is the polaron binding energy (compare
equation (\ref{eio1})) and dominates the
ground state energy (Figure \ref{figgse0.1}).
The second term describes hopping between neighbouring
sites  with the bandwidth reduced by the overlap
of the oscillator ground state centered at $-q_e$
and $+q_e$:
\begin{equation}
J= t \exp \left(- \left( {g \over \omega} \right)^2 \right).
   \label{jeffa1}
\end{equation}
The third term describes the second order process
(of order $t^2/\omega$) where a fermion hops to a neighbouring
site and back again. This term is repulsive because this process 
is not possible if the neighbouring site is occupied.
\begin{equation}
V = {2 J^2 \over \omega} \int_0^{(g/\omega)^2} dx {e^{2x} - 1 \over x}
   \label{veffa1}
\end{equation}
There is an additional term, of second order in $t^2/\omega$,
involving next nearest neighbour inteactions
but it is smaller by a factor 
of order $ 1 / \lambda$ 
and so the 
effective Hamiltonian should accurately describe the
physics in the strong coupling limit,
$ \lambda \gg 1$ or $g^2 \gg t \omega$.
Note that the limit $\omega \to \infty$ corresponds
to free fermions, as pointed out by Hirsch and Fradkin \cite{hir}.

The effective Hamiltonian (\ref{effa1}) is, after
a Jordan Wigner transformation, of the same form
as that of the exactly soluble
 antiferromagnetic XXZ quantum spin chain \cite{frad}.
For convenience we now briefly summarize
some of the known results for this model.
It can be exactly solved by Bethe ansatz \cite{yang,ham}.
The system is metallic for $V<2J$,
i.e., there is no energy gap or long range order
and so it is a Luttinger liquid.
It is in an insulating charge-density-wave state for $V>2J$.
The metal-insulator transition is an infinite order,
i.e. Kosterlitz-Thouless, transition and has been discussed in 
detail by Shankar \cite{sha}.

Define a new variable $\mu$ by
\begin{equation}
\cos \mu = {V \over 2J}
\label{mu}
\end{equation}
where $0 < \mu < \pi/2$.
As $V$ increases from 0 to the transition at $V=2J$,
$\mu$    decreases from $\pi / 2$ to zero.
The velocity of charge excitations is given by \cite{hald}
\begin{equation}
u_\rho=  \pi J {\sin \mu \over \mu} .
\label{luturho}
\end{equation}
As $V$ increases from 0 to the transition at $V=2J$,
the velocity increases from $2J$ to $\pi J$.
However, as the coupling $g$ increases
$J$ rapidly decreases and so $u_\rho$
rapidly decreases (compare Figure \ref{figlut0.1} (a)).
The Luttinger liquid exponent $K_\rho$ is 
\begin{equation}
K_\rho={1 \over 2(1- {\mu \over \pi})} .
\label{lutkr}
\end{equation}
As $V$ increases from 0 to the transition at $V=2J$,
$K_\rho$ decreases from $1$ to $1/2$
(compare Figure \ref{figlut0.1} (b)).
The value $K_\rho=1/2$ is a universal feature
of a Kosterlitz-Thouless transition for one-dimensional
fermions \cite{sha,gia}. 

For $V$ larger than $2J$
define a new variable $\gamma$ by
\begin{equation}
\cosh \gamma = {V \over 2J} .
\label{gamma}
\end{equation}
 The charge-density-wave order parameter (\ref{op1}) is \cite{baxter}
\begin{equation}
m_e= {1 \over 2} \Pi_{m=1}^{\infty}
\tanh^2(m \gamma) .
\label{lutop}
\end{equation}
The coupling dependence is shown in Figure \ref{figgap0.1} (a)
for $t=0.1 \omega$.
The metal-insulator transition is Kosterlitz-Thouless
although it does not appear so on the scale shown.
The energy gap in the insulating phase is
\begin{equation}
\Delta= 2 J \sinh \gamma
\sum_{m= -\infty}^{m=\infty} { (-1)^m \over \cosh(m \gamma)}
\label{lutgap}
\end{equation}
and turns out to be very small
(compare Figure \ref{figgap0.1} (b)).

\subsection{Free fermions ($g=0$)}
\label{secfree}

The fermion states are plane waves with energy dispersion
\begin{equation}
E(k)= -2 t \cos(k) .
\label{fc1}
\end{equation}
These states are occupied for $|k| < k_F \equiv \pi /2$.
Near the Fermi surface at $k= \pm k_F$ we have
$E(k)=  \pm 2 t (k \mp k_F)$ and so the
Fermi velocity is $v_F= 2t$.
The ground state energy per site is
\begin{equation}
\epsilon_\infty = - 2 t \int_{-\pi/2}^{\pi/2}
{d k \over 2 \pi} \cos(k) = -{ 2 t \over \pi} .
\label{fc2}
\end{equation}

\subsection{Adiabatic or mean-field limit ($\omega
\ll te^{-1/\lambda}$)}
\label{secmf}

It is assumed that the fluctuations
of the lattice about its dimerized value can
be neglected and the quantum operator $q_i$
in the Hamiltonian (\ref{ab1}) is replaced 
by its mean value: $q_i \to <q_i>= (-1)^i m_p$.
The fermionic  Hamiltonian can then be diagonalized
by a Bogoliubov transformation and the 
fermionic energies are
\begin{equation}
E(k)= \pm \left(  (2 t \cos(k))^2 + \Delta^2 \right)^{1 \over 2}
\label{fcmf}
\end{equation}
where $\Delta \equiv g (2 M \omega)^{1/2} m_p$ is the 
energy gap at the Fermi surface due to the dimerization.
$\Delta$ is then treated as a variational parameter
and the total energy of the system is minimized
to give the self consistent equation
\begin{equation}
1 =  \lambda t
 \int_{-\pi/2}^{\pi/2} {d k } 
 { 1 \over \left(  (2 t \cos(k))^2 + \Delta^2 \right)^{1 \over 2}} .
\label{fcmf2}
\end{equation}
The system is dimerized for all coupling strengths
and for weak coupling ($\lambda < 1$)
the energy gap is
\begin{equation}
\Delta=8 t  \exp {\left(-1 \over \lambda \right)} .
\label{mfgap}
\end{equation}
The charge-density-wave order parameter is
\begin{equation}
m_e= {\Delta \over 2 \pi \lambda t} 
={4 \over \pi \lambda} \exp {\left(-1 \over \lambda \right)} .
\label{mfop}
\end{equation}
The corrections to 
the mean-field equation (\ref{fcmf2}), to next order in
$\omega \lambda/ \Delta$, were recently calculated
\cite{wu}.


\section{ The Green's Function Monte Carlo Method }
\label{secgfqmc}

At first we tried simulating the model using a discrete
basis of free phonon eigenstates on each site
and employing a ``stochastic truncation'' \cite{pri}
technique appropriate to this basis.
This method gave accurate results for small coupling
$g$, but not at or beyond the metal-insulator transition.
In this region the staggered displacement $m_p$ becomes 
large, corresponding to the presence of highly excited 
states in the free phonon eigenstate basis.
It was thus found more appropriate to use
a continuous ``position space'' basis with variables $\{ q_i \}$,
and use a different Monte Carlo technique as 
described below.

\subsection{Ground state energy}

To simulate the model, we use a Green's Function Monte Carlo (GFMC)
method, as developed by Kalos and collaborators \cite{cep,kal}, and applied
to lattice gauge theory by Chin, Negele and Koonin \cite{chi} and others
\cite{hey,ham2,ham3}. Let us review the method briefly. 

The Hamiltonian for the Holstein model (\ref{ab1})
 can be rescaled to the dimensionless form
\begin{equation}
H= -\tilde t \sum_i \left(c_i^\dagger c_{i+1} + c_{i+1}^\dagger c_i \right)
+ \sum_i p_i^2 + q_i^2 - \tilde g q_i (n_i - {1 \over 2})
\label{gf0}
\end{equation}
where $ \tilde t \equiv 2t/\omega,$
$ \tilde g \equiv 2 \sqrt{2} g/\omega $. 
The imaginary-time Schrodinger equation for the system reads
\begin{equation}
-{\partial \over \partial \tau}
|\Phi(\tau)> = (H - E_T)
|\Phi(\tau)> 
\label{gf1}
\end{equation}
where $E_T$ is a trial energy, representing a constant shift in the zero
of energy.  Evolving this equation for time $\Delta \tau$  yields
\begin{equation}
|\Phi(\tau + \Delta \tau)> = \exp \left( \Delta \tau (E_T - H) \right)
|\Phi(\tau)> .
\label{gf2}
\end{equation}
At large times $\tau$ the ground state will dominate:
\begin{equation}
|\Phi(\tau)> \sim
c_0 \exp \left( -(E_0 - E_T) \tau \right)
|\Phi_0>   \ \ \ \ {\rm as} \ \ \ \ {\tau \to \infty}
\label{gf3}
\end{equation}
where $|\Phi_0>$
is the (time-independent) ground state of $H$
with energy $E_0$.

We shall work in a position-space representation, where the wave
function
\begin{equation}
\Phi(\{q,n\},\tau) 
= <\{q,n\}|\Phi(\tau)> 
\label{gf4}
\end{equation}
and $|\{q,n\}>$  represents an eigenstate of the positions $\{q_i\}$ 
and fermion occupation numbers $\{n_i\}$ at each site. In this representation,
\begin{equation}
H=H_0 + H_1
\label{gf5}
\end{equation}
where 
\begin{equation}
H_1= -\tilde t \sum_i \left(c_i^\dagger c_{i+1} + c_{i+1}^\dagger c_i \right)
\label{gf6}
\end{equation}
is the fermion hopping term, and
\begin{equation}
H_0= - \sum_i {\partial^2 \over \partial q_i^2} + V(\{q,n\})
\label{gf7}
\end{equation}
with
\begin{equation}
 V(\{q,n\})= \sum_i q_i^2 - \tilde g \sum_i q_i (n_i - {1 \over 2})
\label{gf8}
\end{equation}
as the ``potential'' term.

The evolution equation (\ref{gf1}) now has the form of a diffusion equation
in configuration space.  It is assumed that the ground-state wave
function can be chosen  positive everywhere, and it is simulated by
the density distribution of an ensemble of random walkers in
configuration space, whose time evolution mimics that of equation 
(\ref{gf2}).

To obtain good accuracy, one needs to introduce some {\it variational
guidance}, which can be done as follows. Let $|\Psi_T>$
be a trial vector, e.g.,
some variational approximation to the true ground-state eigenvector
    with wave function:
\begin{equation}
\Psi_T(\{q,n\}) 
=<\{q,n\}|\Psi_T> .
\label{gf9}
\end{equation}
Then carry out a similarity transformation
\begin{equation}
|\Phi(\tau)> 
\to |\Phi^\prime(\tau)> 
=\Psi_T 
|\Phi(\tau)> 
\label{gf10}
\end{equation}
\begin{equation}
H \to H^\prime
=\Psi_T H \Psi_T^{-1} 
\label{gf11}
\end{equation}
where the transformation matrix  $\Psi_T$ 
is diagonal in $\{q,n\}$    space, with
diagonal entries 
$\Psi_T(\{q,n\})$.
The modified evolution equation will be
\begin{equation}
|\Phi^\prime(\tau + \Delta \tau)>
 = \exp \left( \Delta \tau (E_T - H^\prime) \right)
|\Phi^\prime(\tau)> 
\label{gf12}
\end{equation}
Let us now separate the fermion hopping term from the rest of the
Hamiltonian, and write for small  
$\Delta \tau$
\begin{equation}
\exp \left( \Delta \tau (E_T - H^\prime) \right)
\simeq 
\exp \left( \Delta \tau (E_T - H_0^\prime) \right)
[1 - \Delta \tau H_1^\prime] + O(\Delta \tau^2)
\label{gf13}
\end{equation}
(All our calculations from here on will only be accurate to
$ O(\Delta \tau) $).

Now $H_0$ transforms to
\begin{eqnarray}
 H_0^\prime
&=&\Psi_T 
[ - \sum_i {\partial^2 \over \partial q_i^2} + V(\{q,n\})]
\Psi_T^{-1} 
\nonumber\\
&=&
 - \sum_i [{\partial^2 \over \partial q_i^2}
 +2 \Psi_T \left({\partial \Psi_T^{-1} \over \partial q_i}\right)
 {\partial \over \partial q_i}
 +  \Psi_T \left({\partial^2 \Psi_T^{-1} \over \partial q_i^2}\right)
] + V(\{q,n\})
\nonumber\\
&=&
\Psi_T^{-1} H_0 \Psi_T 
+ \sum_i [ p_i^2 + 2 i p_i \left( \Psi_T^{-1} {\partial
\Psi_T  \over \partial q_i}\right) ] 
\label{gf14}
\end{eqnarray}
as shown by Chin, Negele, and Koonin \cite{chi} where the operator 
$p_i= -i {\partial \over \partial q_i}$    acts on
everything to the right of it as usual.  Then the matrix element
between position eigenstates corresponding to the  time-step
$\Delta \tau$ at iteration $m$  can be
written \cite{chi}
\begin{eqnarray}
&<&\{q,n\}^{(m+1)}|
\exp \left( \Delta \tau (E_T - H_0^\prime) \right)
|\{q,n\}^{(m)}>
\nonumber\\
&\simeq& 
{1 \over (4 \pi \Delta \tau)^{N/2}}
\exp\left(- {1 \over 4 \Delta \tau}
\sum_i [q_i^{(m+1)} -q_i^{(m)} -2 \Delta \tau 
\Psi_T^{-1} {\partial \Psi_T \over \partial q_i}]^2
-\Delta \tau[\Psi_T^{-1}(H_0\Psi_T) - E_T]\right)
\nonumber\\
 &+& O(\Delta \tau^2)
\label{gf15}
\end{eqnarray}

Representing the wave function $\Phi^\prime$
by a distribution of random walkers
in position space, the Monte Carlo simulation proceeds as follows.
Each iteration corresponds to a time step $\Delta \tau$, and involves a sweep
through each site in turn.  The first term in the exponential (\ref{gf15})
is simulated by a displacement of each position variable
\begin{equation}
\Delta q_i = 2 \Delta \tau 
\Psi_T^{-1} {\partial \Psi_T \over \partial q_i}
+ \chi
\label{gf16}
\end{equation}
where $\chi$ is randomly chosen from a Gaussian distribution with standard
deviation $\sqrt{2 \Delta \tau}$. 
 The first term in (\ref{gf16}) is the ``drift'' term, and the 
second is the ``diffusion'' term. The second term in the exponential
(\ref{gf15}) is simulated by multiplying the "weight" of each walker by an
equivalent amount.

We also need to simulate the fermion hopping term:
\begin{eqnarray}
&<&\{q,n\}^{(m+1)}|
[1 - \Delta \tau H_1^\prime]
|\{q,n\}^{(m)}>
\nonumber \\
&=&
{\Psi_T(\{q,n\}^{(m+1)}) \over
\Psi_T(\{q,n\}^{(m)})}       
<\{q,n\}^{(m+1)}|
[1 + \tilde t \Delta \tau
 \sum_i \left(c_i^\dagger c_{i+1} + c_{i+1}^\dagger c_i \right)]
|\{q,n\}^{(m)}> 
\label{gf17}
\end{eqnarray}
The factor in front produces a ``reweighting'' of the walkers in the
ensemble; while the hopping term itself produces new configurations
on walkers with different fermion occupation numbers.

At the end of each iteration, the trial energy $E_T$ is adjusted to
compensate for any change in the total weight of all walkers in the
ensemble; and a ``branching'' process is carried out, so that walkers
with weight greater than (say) 2 are split into two new walkers, while
any two walkers with weight less than (say) 0.5 are combined into one;
chosen randomly according to weight from the originals. This procedure
of ``Runge smoothing'' \cite{run}  maximises statistical accuracy by keeping
the weights of all walkers within fixed bounds, while minimizing
any fluctuations in the total weight due to the branching process.
When equilibrium is reached after many  sweeps through the lattice,
the average value of the trial energy $E_T$  will give an estimate of the
ground-state energy $E_0$, from equation (\ref{gf3}); and the density of the 
ensemble in configuration space will be proportional to  
$\Phi_0 \Psi_T$.
Various corrections due to the finite time interval 
$ \Delta \tau$   have been
ignored in this discussion, and the limit 
$ \Delta \tau \to 0$ must be taken in
some fashion to eliminate such corrections.

As a trial wave function, we choose a Gaussian, displaced by an amount
$q_0$ at each
site depending whether the site is occupied or unoccupied:
\begin{equation}
\Psi_T(\{q,n\})=\exp \left[-c \sum_i(q_i - 2 q_0(n_i-{1 \over 2}))^2
 \right]
\label{gf18}
\end{equation}
where $c$ and $q_0$   are variational parameters.
 Then the local ``trial energy''
\begin{eqnarray}
E_L(\{q,n\}) \equiv
\Psi_T^{-1} H_0 \Psi_T =
 \sum_i[q_i^2 - \tilde g q_i(n_i-{1 \over 2})]
 -\sum_i[4c^2(q_i - 2 q_0(n_i-{1 \over 2}))^2 -2c]
\label{gf19}
\end{eqnarray}
and the ``drift'' term is
\begin{equation}
2\Psi_T^{-1} {\partial \Psi_T \over \partial q_i}
= -4c(q_i - 2 q_0(n_i-{1 \over 2}))
\label{gf20}
\end{equation}
while the ``reweighting factor'' in equation (\ref{gf17}) is
\begin{equation}
{\Psi_T(\{q,n\}^{(m+1)}) \over
\Psi_T(\{q,n\}^{(m)})}       
=\exp \left[4 c q_0 \sum_i q_i (n_i^{(m+1)}
 - n_i^{(m)}) \right] 
\label{gf21}
\end{equation}

If the choice of trial function is a good one, and $E_T$   is adjusted to
approximately equal $E_0$  , then we will have
\begin{equation}
E_L \simeq E_T \simeq E_0
\label{gf22}
\end{equation}
so that the weight of each walker changes very little at each time
step, according to equation (\ref{gf17}), so that the fluctuations in the
weights are small, and consequently the accuracy of the calculation
is maximised.

\subsection
{Expectation Values}

Ground-state expectation values can also be measured, using a 
``secondary amplitude'' technique discussed by Hamer {\it et al.} 
\cite{pri,ham2,ham3}.
Let  $<Q>_0$
   be the expectation value to be measured, where we assume
the operator $Q$  is diagonal in the $\{q,n\}$ representation.  Use 
$Q$ as a perturbation to the Hamiltonian:
\begin{equation}
H'=H + x Q.                 
\label{gf23}
\end{equation}
Let $E'_0(x)$ denote the ground state expectation value of
this Hamiltonian.
By the Hellmann-Feynman theorem, the required expectation
value is given by
\begin{equation}
<Q>_0= { d E'_0 \over d x} \bigg|_{x=0} .
\label{gf24}
\end{equation}
Taylor expand the eigenvector and eigenvalue
\begin{equation}
|\Phi(\tau,x)> =
|\Phi^0(\tau)> +
x |\Phi^1(\tau)> + O(x^2) 
\label{gf25}
\end{equation}
\begin{equation}
E'_0(x)= E_0 + x E^1 + O(x^2) 
\label{gf26}
\end{equation}
substitute in the evolution equation (\ref{gf2}) (ignoring any variational
guidance for the time being), and equate powers of $x$ to obtain:
\begin{equation}
|\Phi^0(\tau + \Delta \tau)> = \exp \left( \Delta \tau (E_T - H) \right)
|\Phi^0(\tau)> 
\label{gf27}
\end{equation}
and
\begin{equation}
|\Phi^1(\tau + \Delta \tau)> = \exp \left( \Delta \tau (E_T - H) \right)
|\Phi^1(\tau)> 
+ \Delta \tau (E^1 - Q)|\Phi^0(\tau)> .
\label{gf28}
\end{equation}
Equation (\ref{gf27}) is just the original evolution equation
(\ref{gf2}) for the
unperturbed system.  Equation (\ref{gf28}) is an evolution equation of
similar structure for the first-order wave function
$|\Phi^1>$.   It is
simulated by giving a ``secondary'' weight to each walker in the
ensemble, and evolving it according to (\ref{gf28}); while a secondary
trial energy $E_T'$  is used to estimate $E^1$  ,  and is adjusted after
each iteration to compensate for any change in the total of all
secondary weights.  At equilibrium, the average value of $E_T'$  gives
an estimate of $E^1$, which is equivalent to   $<Q>_0$
  by equation (\ref{gf24}).

\twocolumn
\section
{Results and Discussion}
\label{secres}

GFMC runs were performed for a range of different couplings $g/\omega$
 at hopping
parameter values $t=0.1\omega, \omega \ {\rm and} \  10 \omega$,
 for lattice sizes of 2, 4, 6, 8, and 16 sites.
In each case, the variational parameters $c$ and $q_0$ 
(compare equation (\ref{gf18})) were adjusted to their
optimum values by a series of trial runs. Production runs typically
employed an ensemble of 2000 walkers for 20,000 iterations.  The first
2000 iterations were discarded to allow for equilibrium, and the remainder
were averaged over blocks of 1024 iterations before estimating the
error to minimize correlation effects.
Calculations were performed on a cluster of six
HP735 workstations. A typical run for 16 sites
took 1-2 hours of CPU time.
Two different time steps were used
in each case, namely $\Delta \tau =  0.005 $ and 0.01 at
 $t=0.1 \omega$, $\Delta \tau  = 0.0005$ and 0.001 at
$t= \omega$, and $\Delta \tau = 0.001$ and 0.002 at
 $t=10 \omega$.
  The results were then linearly
extrapolated to $\Delta \tau  =0$.

The quantities measured were the ground-state energy (in the half-filled
sector), $E_0(N)$, the energy gap (to the ``one hole'' sector with one fewer
 fermions), $E_{-1}(N)-E_0(N)$,
and ground-state expectation values for the mean displacement 
$<q_i(2n_i-1)>$,
the mean square displacement $<q_i^2>$,
 and two correlated fermion expectation
values,  $<n_i n_{i+ N/2}>$ and $<n_i n_{i-1 + N/2}>$,
 where $N$ is the lattice size.
The difference between these last two values provides an estimate of the
amount of ``staggering,'' or dimerization, in the fermion occupation 
numbers. (Compare equation (\ref{me12}) below).
A sample of results is shown in Table \ref{table1}.

The charge velocity $u_\rho$
 was extracted from a finite size scaling plot of
the ground state energy per site $E_0(N)/N$  versus 
$1/N^2$  (compare Figure \ref{fige0}).
According to equation (\ref{e0})  this should be a straight line for large
$N$. To allow for the small curvature of our plots, because we used
only moderately large system sizes
($N$ = 2, 4, 6, 8, and 16 sites), the  data was fitted to
\begin{equation}
{E_0(N) \over N}= \epsilon_\infty - {\pi u_\rho \over 6 N^2}
 + {a \over N^4} .
\label{e02}
\end{equation}
The correction-to-scaling term $O(N^{-4})$ matches
that predicted to hold
for the XXZ model \cite{yang,ham}, at least
for weak interactions (i.e., small $\mu$).
For stronger interactions the exponent is interaction
dependent. At   the
metal-insulator transition the correction-to-scaling term is
$O((N \ln N)^{-2})$.  However, we found that
using such a form did not improve the quality of the
least square fits near the transition.

For free fermions ($g=0$) the values of  
$\epsilon_\infty$ and  $u_\rho$ extracted from the fits
were found to agree well with the known analytic results 
$\epsilon_\infty = -2t/\pi $ and  $u_\rho = 2t$ (Table \ref{table2}).
For $t=0.1\omega$ the dependence of the ground state energy on
the coupling is in good agreement with small polaron theory
(Figure \ref{figgse0.1}).

The energy gap $\Delta$  of the infinite system
 was extracted from finite size
scaling plots of the hole energy $E_{-1}(N)-E_0(N) $ 
 versus $1/N$ (compare Figure \ref{fige1}).  To allow for the
 corrections to scaling this was fitted to
\begin{equation}
E_{-1}(N)-E_0(N) = \Delta + { \pi u_\rho  \over 2 K_\rho  }{1 \over N} 
+ {b \over N^3} .
\label{e12}
\end{equation}
Again,the higher order term $O(N^{-3})$ was chosen to be consistent with
known results for free fermions and the XXZ model with
weak interactions. To extract $K_\rho$  we need to use the value
of $u_\rho$  extracted earlier.
(Strictly speaking equation (\ref{e12}) is only valid
when $\Delta=0$ but we use $\Delta$ as a parameter in our
fits to check that we are in the critical regime.
Also, the derivation of equation (\ref{e12})
requires particle-hole symmetry, i.e.,
$E_{-1}(N)=E_{+1}(N)$. We checked for several parameter
values  that the Monte Carlo
results were consistent with this.)

  Figure \ref{figlut0.1} (a) shows the dependence of the charge
velocity  $u_\rho$ on the fermion-phonon coupling $g$
for $t=0.1 \omega$.
  The results are in good
agreement with equations
(\ref{jeffa1}) and (\ref{luturho}) (solid line in Figure \ref{figlut0.1}(a)).
The charge velocity is significantly reduced by
polaronic band narrowing.
  The correlation exponent $K_\rho$ is shown in
Figure \ref{figlut0.1} (b) as a function of the
fermion phonon coupling.
The dependence of $K_\rho$  on the coupling
is consistent with the metallic phase being a
Luttinger liquid.
 The fact $K_\rho < 1$ indicates repulsive 
interactions in the Luttinger liquid.
$K_\rho $
 is not plotted for $g > 1.5 \omega$
 because the relative error is very large.
This is because its determination depends on the
value of $u_\rho$   which has a very large
relative error for $g > 1.5 \omega$
(see Figure \ref{figlut0.1} (a)).  For $t = 0.1 \omega$, equations
(\ref{jeffa1}) and (\ref{veffa1}),
 for the small polaronic model together with the criterion
$V=2J$ can be used to determine that the transition from the Luttinger
liquid to the insulating phase occurs when $g = 2.075 \omega$.

  The charge-density-wave order parameter $m_e$, defined by (\ref{op1}),
must  be zero for any 
finite size system. However, in the dimerized phase we also have for
$j$ large
\begin{equation}
<n_i n_{i+j}> = {1 \over 4} + (-1)^j m_e^2
\label{me12}
\end{equation}
and so
\begin{equation}
m_e^2 = {1 \over 2}| <n_i n_{i+N/2}> - <n_i n_{i-1+N/2}>| .
\label{me13}
\end{equation}
This equation was used to determine $m_e^2$
from the results for $N=16$ sites.

Figure \ref{figgap0.1} (a) shows the coupling constant dependence
of $m_e^2$ for $t = 0.1 \omega$.
The quantum Monte Carlo data suggests there is a transition near
$g=1.8 \omega$.
This is consistent with the small polaron theory prediction of
 $g = 2.075 \omega$
since the latter theory is only valid to order 
$1 / \lambda \sim  \pi t \omega / g^2$ , i.e., about 10\%.  Figure
\ref{figgap0.1} (b) shows the energy gap as a function of coupling.  It is not
possible to detect the transition in the energy gap data.  Small polaron
theory predicts an  energy gap smaller than typical uncertainties in the
Monte Carlo data.

Figure \ref{figq0} shows the coupling strength dependence of
the mean lattice displacement
$<q_i(2n_i-1)>$
and the mean square lattice displacement $<q_i^2>$    
for $t=0.1\omega$ and a system of 16 sites.
The results are very close to those anticipated for
localized fermions
(compare equations (\ref{qo1}) and (\ref{q21})).
For $t=\omega$ and $t=10\omega$ the mean lattice displacement
was also non-zero, i.e., the ground state was polaronic for all
couplings, although the magnitude of the displacement
decreased significantly with increasing $t$. Similar
trends are seen for the Holstein model on two sites \cite{son}.

For $t = \omega$ the charge velocity is again reduced by polaronic effects
(Figure \ref{figlut1} (a)) but not by as much as for $t = 0.1 \omega$.
 The interactions in the
Luttinger liquid are now attractive ($K_\rho > 1$).  Both the order parameter
and the energy gap show a transition to the insulating phase near
$g = 1.7 \omega$ (Figure \ref{figgap1}).
Clearly our results are inconsistent with the universal
value $K_\rho=1/2$ expected for a Kosterlitz-Thouless transition
\cite{sha,gia}.

For $t = 10 \omega$ the charge velocity decreases 
by less than ten per cent 
 with increasing 
$g/\omega$ (Figure \ref{figlut10}).
This is in contrast to the cases of
$t=0.1\omega$ and $t = \omega$ for which the
charge velocity decreases by about an order
of magnitude.
The correlation function exponent $K_\rho$ 
increases by about fifty per cent.
As for $t = \omega$, $K_\rho \neq 1/2$ at the
transition and so the transition cannot
be a Kosterlitz-Thouless transition.
 The order parameter $m_e$  and energy gap  $\Delta$
  become nonzero about $g=3.5 \omega$ (Figure
\ref{figgap10}).
  This is quite different from what is anticipated by Hirsch and
Fradkin \cite{hir}.  They performed simulations from
$t = 0.5 \omega$ up to $t = 3.1 \omega$.
They found a smooth decrease in the critical value of $\lambda_c
=g_c^2/\pi \omega t$ with increasing $t / \omega$,
and anticipated a smooth crossover to
$\lambda_c=0$ for $\omega=0$.  Extrapolating
their results to $t=10 \omega$ gives  $\lambda_c
\sim   0.01$  and $g_c  \sim 0.6 \omega$ compared
to our value of $g_c \simeq 3.5 \omega$.  Note     that the
ratio  of the energy gap to its mean field value
 (Figure \ref{figgap10} (b)) is much smaller than
the ratio of the charge-density-wave 
order parameter to its mean field value.
This is consistent with work showing 
that the zero point motion of the lattice
can reduce the magnitude of
the order parameter by a small amount but
produce a substantial subgap tail in
the fermionic density of states \cite{mck}.
(For example, results
on the continuum version of the SSH model
shown in Figures 1 and 3 of Reference \cite{mck}
show that for one set of parameter values the energy gap
can be about 60 \% of the mean-field value
while the order parameter is only reduced by
about 5 \%).

The phase boundary as a function of $t/\omega$ and
$g/\omega$   between the metallic and insulating
phases is shown in Figure \ref{figphased}.
The solid curve is the boundary predicted by
small polaron theory and the XXZ model (Section \ref{secpol}).
This curve is only shown for $t < \omega$
since this model is only valid in the strong coupling limit
($t \ll g^2/\omega$).
The crosses are the boundary points
deduced from Figures \ref{figgap0.1},
\ref{figgap1}, and \ref{figgap10}.
It should be stressed that there is some ambiguity in
determining the phase boundary.
According to
mean-field theory the transition occurs at $g=0$
but the solid curves in Figures \ref{figgap1} and \ref{figgap10}
suggest that the transition is actually only detectable at
$g \sim 0.6 \omega$ and $g \sim 2 \omega$, respectively.
On the other hand, for $t = 0.1 \omega$
small polaron theory and the XXZ model
predict a Kosterlitz-Thouless
transition at $g=2.075\omega$ and the solid curve
in Figure \ref{figgap0.1} (a) shows there is very little
ambiguity associated with this transition point.
 For comparison the boundary points
found by Hirsch and Fradkin \cite{hir} (Figure 11 in their
paper) are also shown.
For $t=\omega$ there is a discrepancy between our results and
theirs: they observe the transition at smaller coupling than we 
do. We have no explanation for this discrepancy.

\section{Conclusions}
\label{secconc}

We have shown that the Green's function Monte Carlo technique can be
successfully used to investigate a one dimensional fermion-phonon
model.  As far as we are aware this is the first application of this
technique to this important class of models.  The results were of
sufficiently high precision that a finite size scaling analysis of
the results could be performed.  For the case of free fermions ($g=0$)
and the strong coupling limit ($g^2 >>t \omega$) our results agree with known
analytic results.

Our results are consistent with the following
physical  picture of the Holstein
model of spinless fermions at half-filling.  For sufficiently weak
 coupling the system is in a metallic, i.e., gapless phase, with the
properties of a Luttinger liquid, i.e., the exponents associated with
the decay of correlation functions depend on the coupling strength.
The fermions are polaronic, i.e., there is a finite phonon displacement
$q_e$ associated with the occupation of a site by a fermion and the velocity
of excitations, $u_\rho$, is reduced below the free electron value 
$2t $.  As the coupling $g$
increases and $t/\omega$
 decreases $q_e$ increases and $u_\rho$ (which is a measure
of the polaronic band width) decreases.  Qualitatively
similar behaviour is seen for the two-site Holstein model
\cite{son}.  In the anti-adiabatic limit ($t \ll \omega$)
 the effective interaction between polarons is
repulsive and for strong coupling
the Holstein model can be mapped onto the XXZ
 antiferromagnetic spin chain (Section \ref{secpol}).  However, as $t / \omega$
 increases the effective interaction becomes attractive.
  This is indicated by a
change in the value of the stiffness constant $K_\rho$
 from values
less than one to values larger than one.
When the fermion-phonon coupling is
 sufficiently large the system undergoes a transition
to an insulating phase, i.e., an energy gap opens at the Fermi surface.
There is long-range charge-density-wave order and a dimerization of the
phonons in this phase.  Our results for $t=\omega$ and $t=10\omega$
are inconsistent
with the metal-insulator transition being infinite order.
On the other hand we  do not
see any evidence of the first-order transition suggested by Zheng,
Feinberg, and Avignon\cite{zhe}
 and by Wu, Huang, and Sun\cite{wu} for certain parameter
values.

This work suggests several possible future investigations which we plan
to  pursue:
(a)  The adiabatic region of the phase diagram
 ($t \gg \omega$), in which we found
a larger region of the metallic phase than anticipated by Hirsch and
Fradkin, needs to be investigated further.
(b)  The relative importance of superconducting and
 charge-density-wave correlations should be
 investigated in the region of the metallic
phase for which the effective interactions are attractive.
(c)  Alternative variational wavefunctions, such as the double Gaussian
proposed by Shore and Sander\cite{sho}, could be used instead of the single
Gaussian (\ref{gf18})  used for the variational guidance.
Finally, we plan to use this method to investigate the Holstein model
with spin, and away from half-filling, as well
as the Su-Schrieffer-Heeger model,
and the spin-Peierls problem.

\narrowtext
\acknowledgements

Work at UNSW was supported by the Australian Research Council.
We thank J. Voit for a critical reading of the manuscript
and many helpful suggestions.
We thank M. Gulacsi, J. Oitmaa, and R. Singh for helpful discussions.
Computer time was provided by the Centre for Advanced Numerical
Computation in Engineering and Science (CANCES) at UNSW.


\twocolumn
\centerline{\epsfxsize=7.0cm \epsfbox{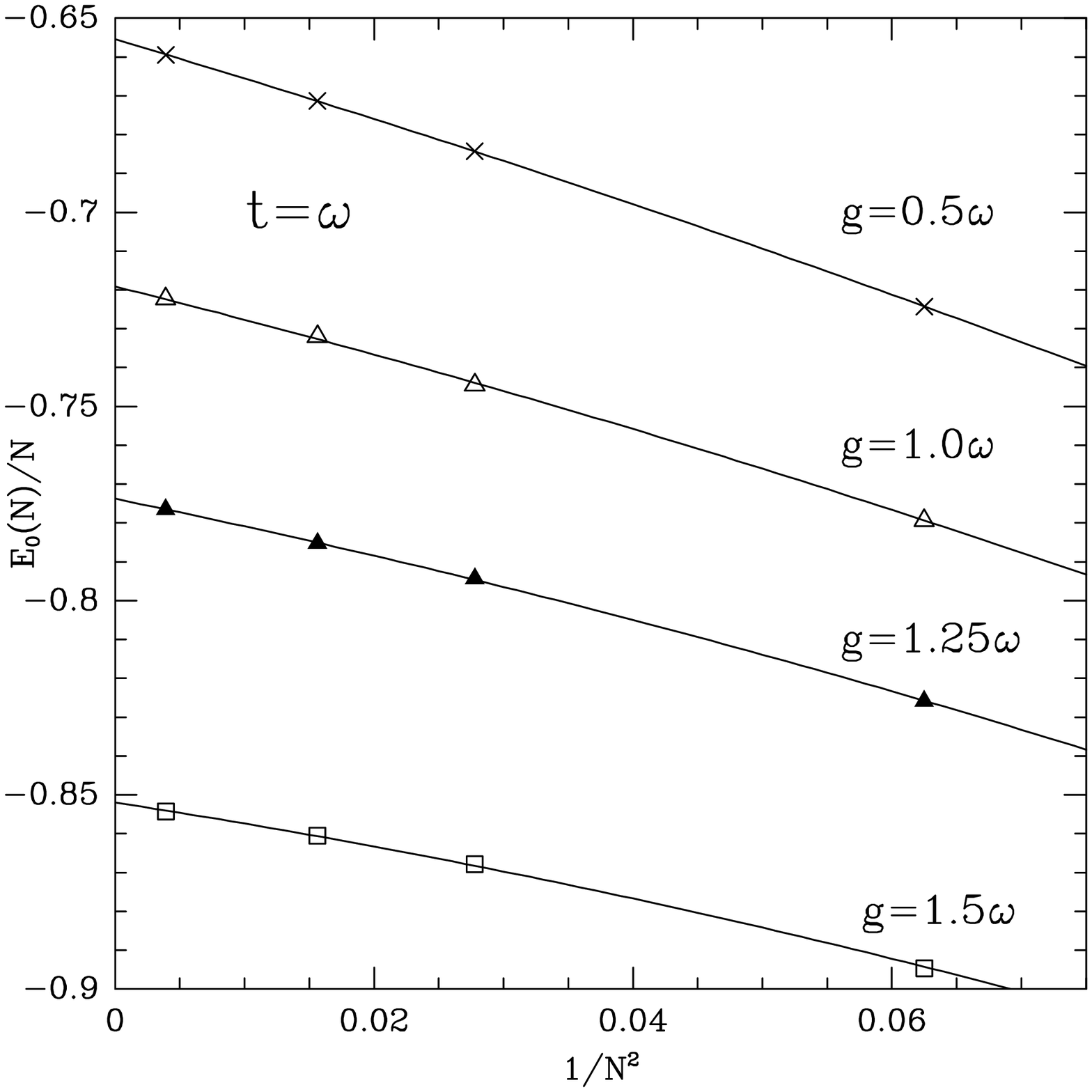}}
\begin{figure}
\caption{
Finite-size scaling of the ground state energy $E_0(N)$
for different values of the fermion-phonon coupling $g$.
The data shown are for $N=$ 4, 6, 8, and 16
lattice sites. All data are for
a fermion hopping parameter $t$ equal to the
phonon frequency $\omega$ and for
a half-filled band (i.e., one fermion per two sites).
All energies are in units of $\omega$.
If the system is critical for a particular $g$ value  then 
the data for $N$ large  should lie on a straight line
(see equation (\protect\ref{e0})).
The lines are least square fits to a parabola (see text).
 The errors in the
Monte Carlo data are smaller than the symbol sizes.
\label{fige0}}
\end{figure}

\centerline{\epsfxsize=7.0cm \epsfbox{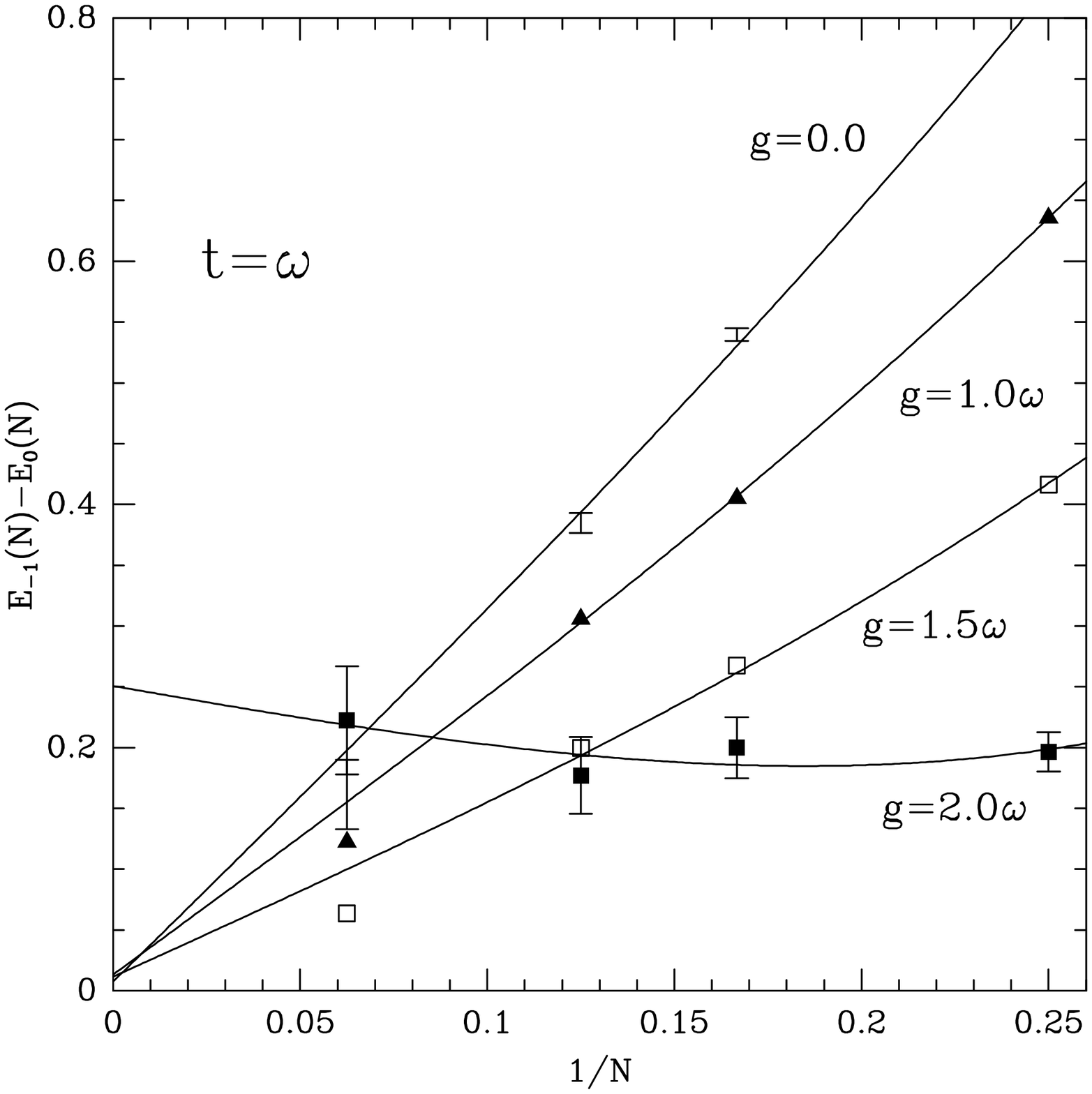}}
\begin{figure}
\caption{
Finite-size scaling of the hole energy 
for different values of the fermion-phonon coupling $g$.
The data shown are for $N=$ 4, 6, 8, and 16
lattice sites. $E_0(N)$ is the ground
state energy of a system of $N/2$ fermions
and $E_{-1}(N)$ is the ground
state energy of a system of $N/2-1$ fermions.
All data are for
a fermion hopping parameter $t$ equal to the
phonon frequency $\omega$.
All energies are in units of $\omega$.
If the system is critical then 
for that $g$ value the data for large $N$ should lie on a straight line
through the origin
(see equation (\protect\ref{e1})).
The lines are least square fits to a cubic (see text).
\label{fige1}}
\end{figure}

\centerline{\epsfxsize=7.0cm \epsfbox{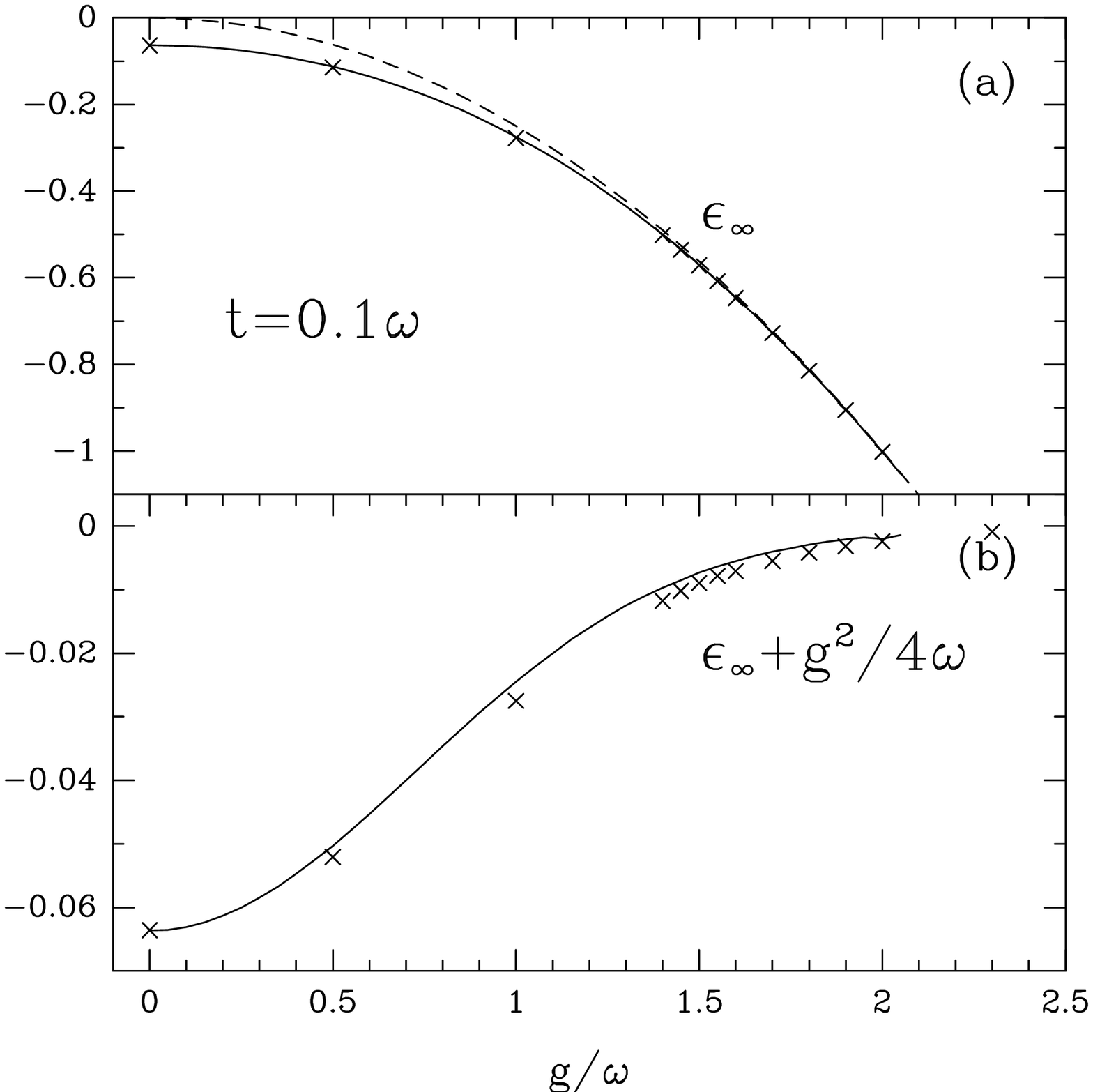}}
\vspace {-0.3 cm}
\begin{figure}
\caption{
Dependence of the ground state energy  per site
$\epsilon_\infty$ on the fermion-phonon coupling $g$ for $t=0.1\omega$.
The solid lines are the predictions of the
small polaron model (Section \protect\ref{secpol}).
The error bars are smaller than the symbol size.
All energies are in units of $\omega$.
(a) 
$\epsilon_\infty$ is deduced from the intercept of the finite-size scaling
plot of the ground state energy (compare Figure
 \protect\ref{fige0}).
The dashed line is the polaron binding energy $-g^2/4 \omega$
and clearly represents almost all of the ground state energy.
(b)
The polaron binding energy has been subtracted from
the ground state energy to make a more accurate comparison with
small polaron theory.
\label{figgse0.1}}
\end{figure}

\vspace {-0.6 cm}
\centerline{\epsfxsize=7.0cm \epsfbox{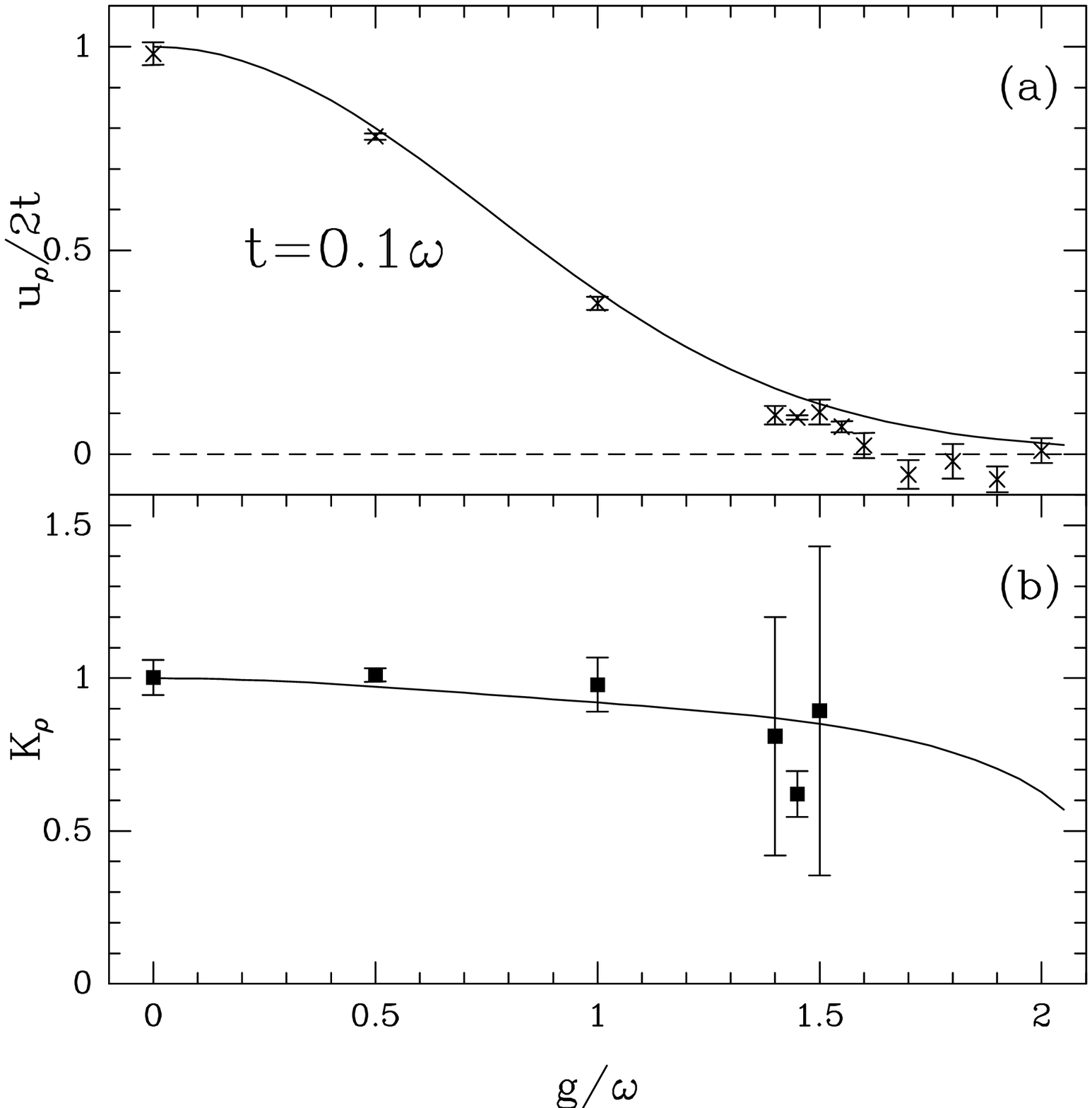}}
\vspace {-0.4 cm}
\begin{figure}
\caption{
Dependence of Luttinger liquid parameters
 on the fermion-phonon coupling $g$ for $t=0.1\omega$.
The solid lines are the predictions of the
small polaron model (Section \protect\ref{secpol}).
(a) The velocity of charge excitations
$u_\rho$ is deduced from the slope of the finite-size scaling
plot of the ground state energy (compare Figure \protect\ref{fige0})
and is normalized by the free fermion value $2t$.
The decrease of 
 $u_\rho$ with increasing $g $ is due to the narrowing of
the bandwidth by polaronic effects.
(b) The correlation function exponent
$K_\rho$ is deduced from the ratio of the slopes
 of the finite-size scaling plots in Figures
\protect\ref{fige0} and \protect\ref{fige1}
(see equation (\protect\ref{e1})).
The error bars are based on the uncertainties
in the least-squares fits to the finite size scaling data
(compare figures \protect\ref{fige0} and \protect\ref{fige1}).
\label{figlut0.1}}
\end{figure}

\centerline{\epsfxsize=7.0cm \epsfbox{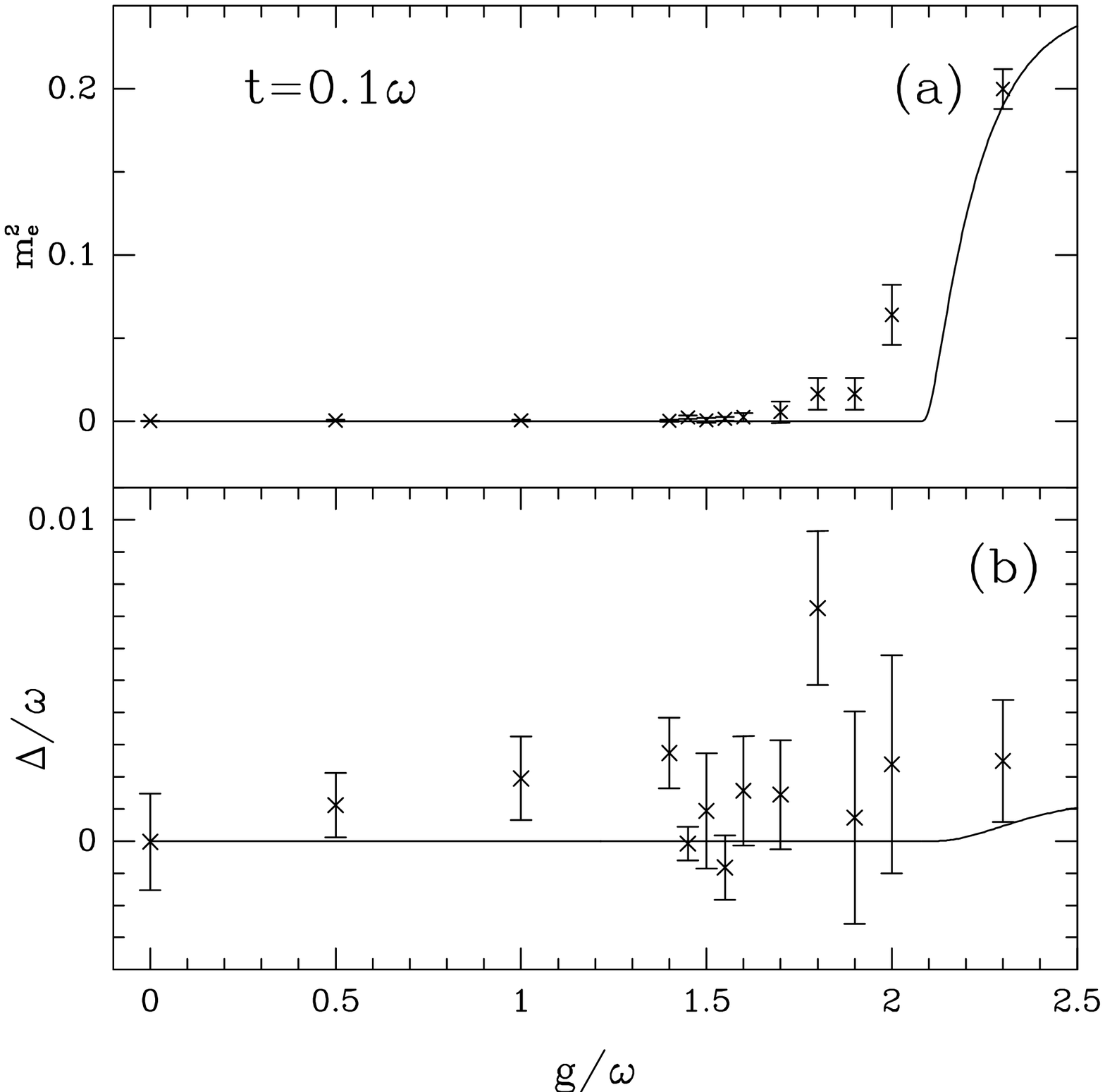}}
\begin{figure}
\caption{
Dependence on the fermion-phonon coupling $g$ of
(a) the square of the charge-density-wave order parameter $m_e$
and (b) the energy gap $\Delta$.
The solid lines are the predictions of the
small polaron model (Section \protect\ref{secpol}).
It predicts an 
infinite order transition at $g=2.075 \omega$.
$m_e^2$ was deduced from equation (\protect\ref{me13})
for a system of 16 sites.
The energy gap was deduced from the $N=\infty$ extrapolation
of the finite size scaling plot of the hole energy
(compare Figure \protect\ref{fige1}).
Both the order  parameter and the energy gap become non-zero
for $g >1.8$ marking the transition into the insulating phase.
\label{figgap0.1}}
\end{figure}

\centerline{\epsfxsize=7.0cm \epsfbox{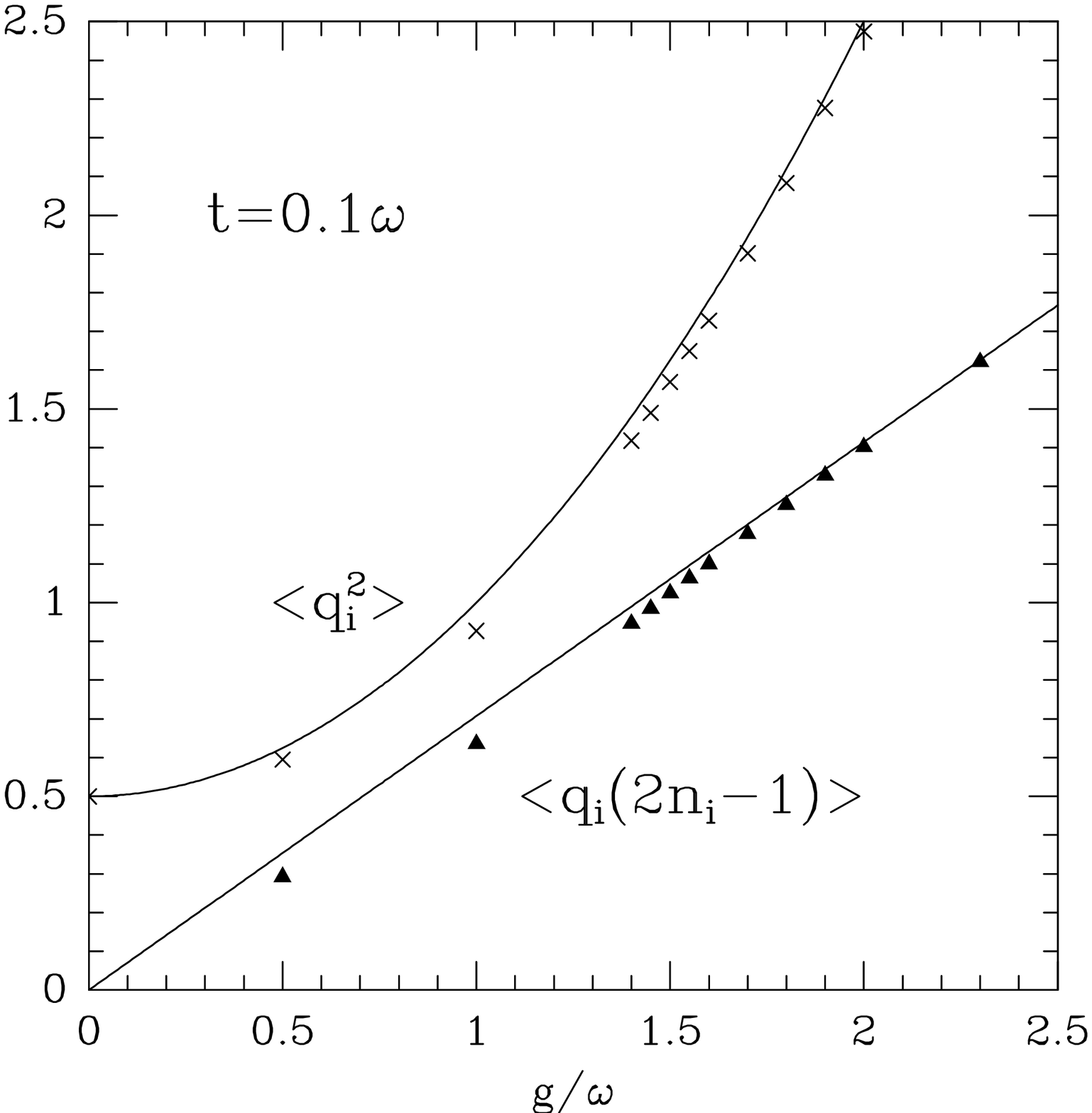}}
\begin{figure}
\caption{
Dependence on the fermion-phonon coupling $g$ of
the mean lattice displacement
$<q_i(2n_i-1)>$
and the mean square lattice displacement $<q_i^2>$    
for $t=0.1\omega$ and a system of 16 sites.
The displacements are in units of $(M \omega)^{-1/2}$.
The solid lines are the predictions for localized fermions
(Section \protect\ref{secloc}).
\label{figq0}}
\end{figure}

\centerline{\epsfxsize=7.0cm \epsfbox{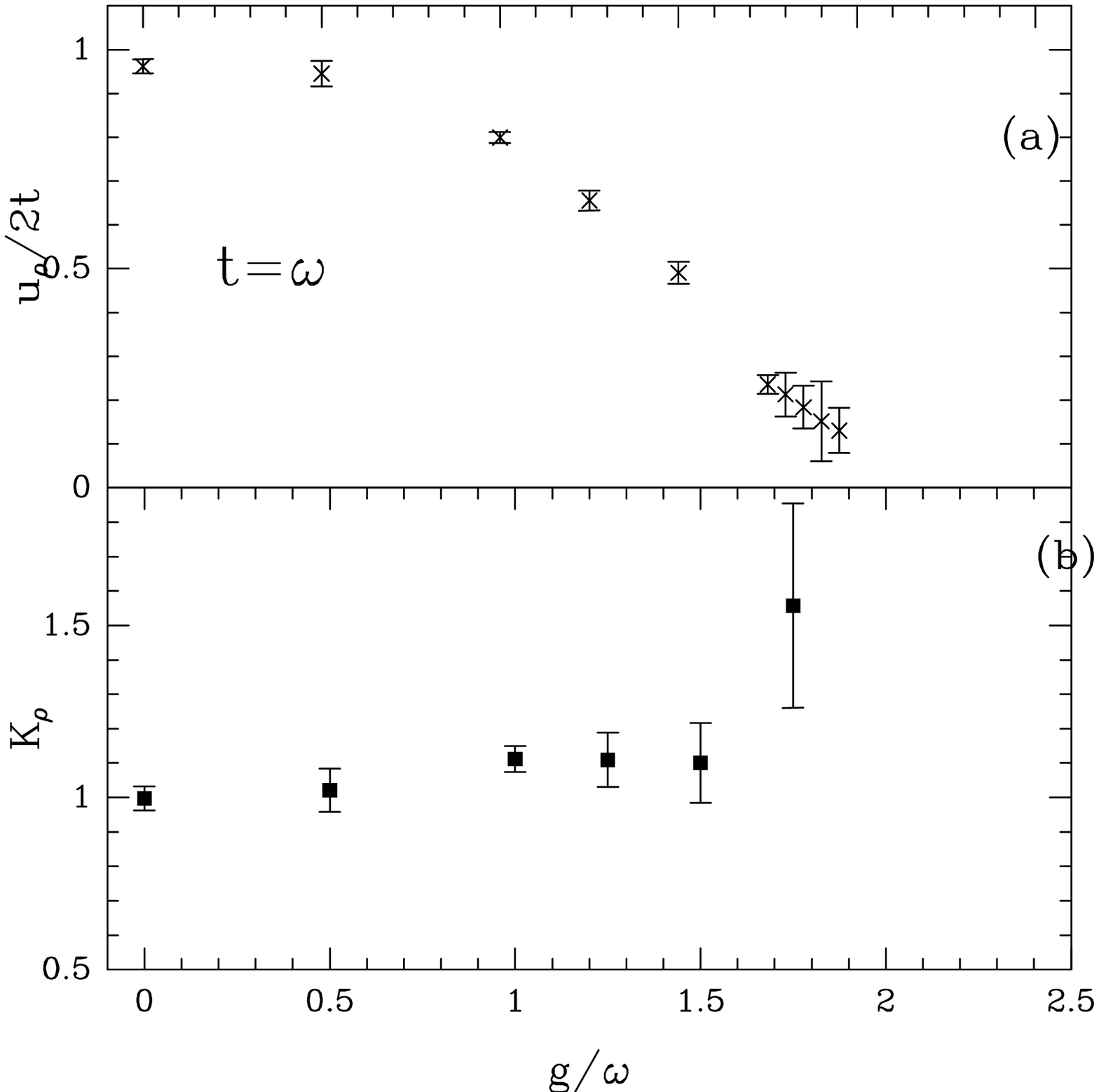}}
\begin{figure}
\caption{
Same as Figure \protect\ref{figlut0.1} but with $t=\omega$.
The fact that $K_\rho$ depends on $g$ and is larger than one
is consistent with the metallic phase being a 
Luttinger liquid with attractive interactions.
If the metal-insulator transition was Kosterlitz-Thouless
$K_\rho$ would equal 0.5 at the transition.
\label{figlut1}}
\end{figure}

\centerline{\epsfxsize=7.0cm \epsfbox{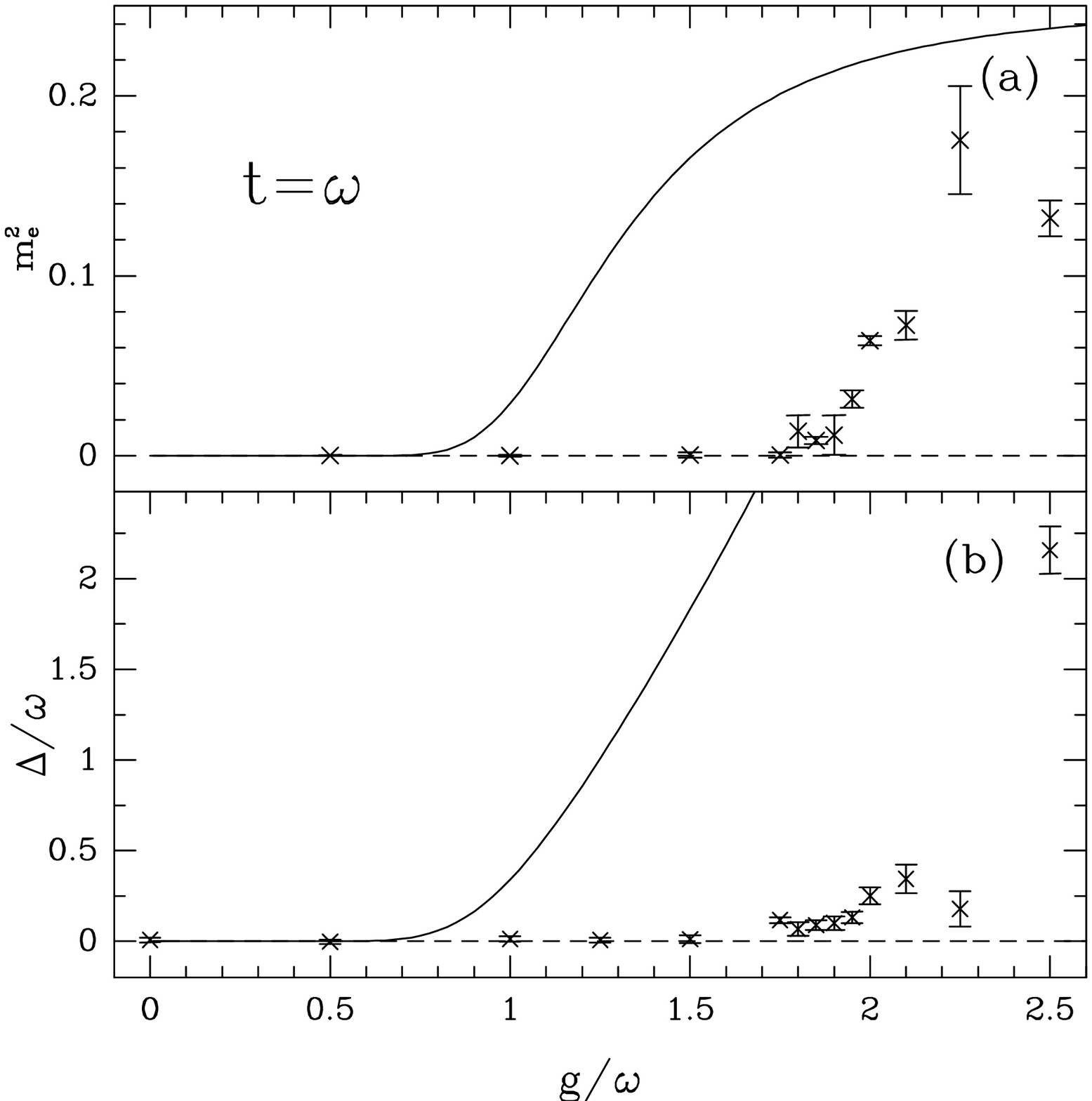}}
\begin{figure}
\caption{
Same as Figure \protect\ref{figgap0.1} but with $t=\omega$.
The solid curves are the predictions of mean field theory
(Section \protect\ref{secmf}).
\label{figgap1}}
\end{figure}

\centerline{\epsfxsize=7.0cm \epsfbox{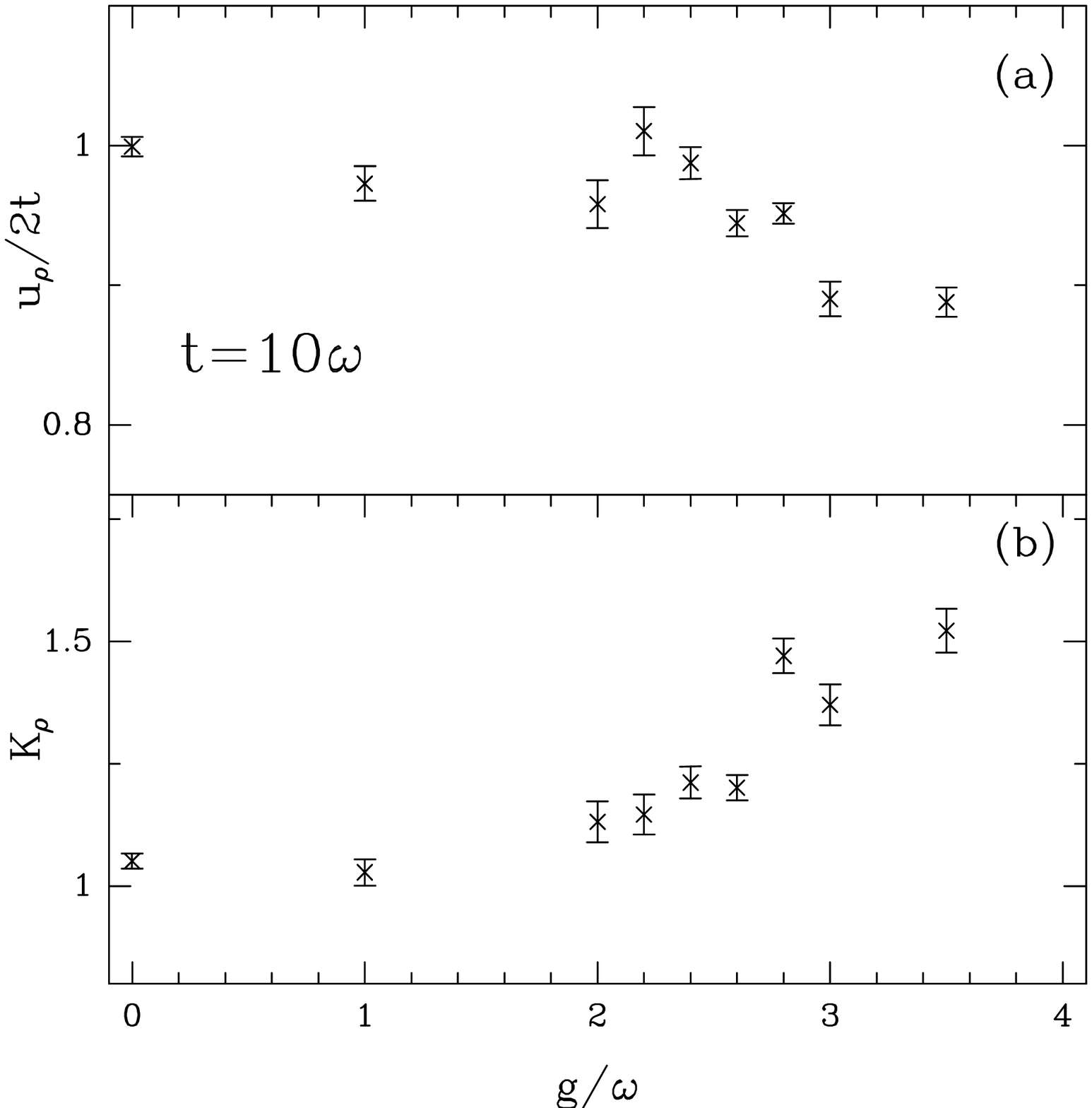}}
\begin{figure}
\caption{
Same as Figure \protect\ref{figlut0.1} but with $t=10\omega$.
Note that the vertical scale is expanded compared to
Figures \protect\ref{figlut0.1}  and \protect\ref{figlut1}.
Since $K_\rho \neq 0.5$ near the transition
the transition is not Kosterlitz-Thouless.
\label{figlut10}}
\end{figure}

\centerline{\epsfxsize=7.0cm \epsfbox{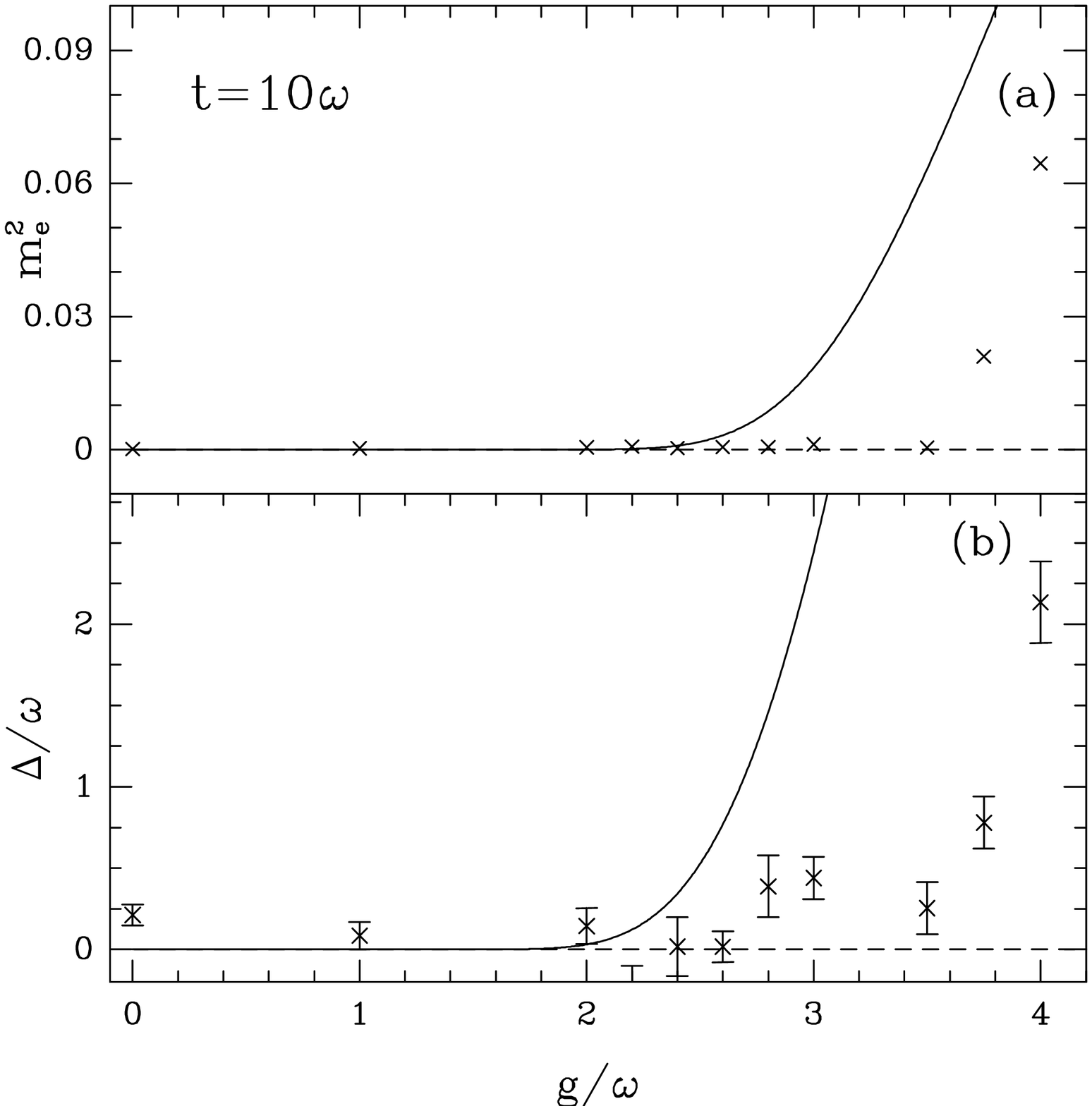}}
\begin{figure}
\caption{
Same as Figure \protect\ref{figgap0.1} but with $t=10\omega$.
The solid curves are the predictions of mean field theory
(Section \protect\ref{secmf}).
\label{figgap10}}
\end{figure}

\centerline{\epsfxsize=7.0cm \epsfbox{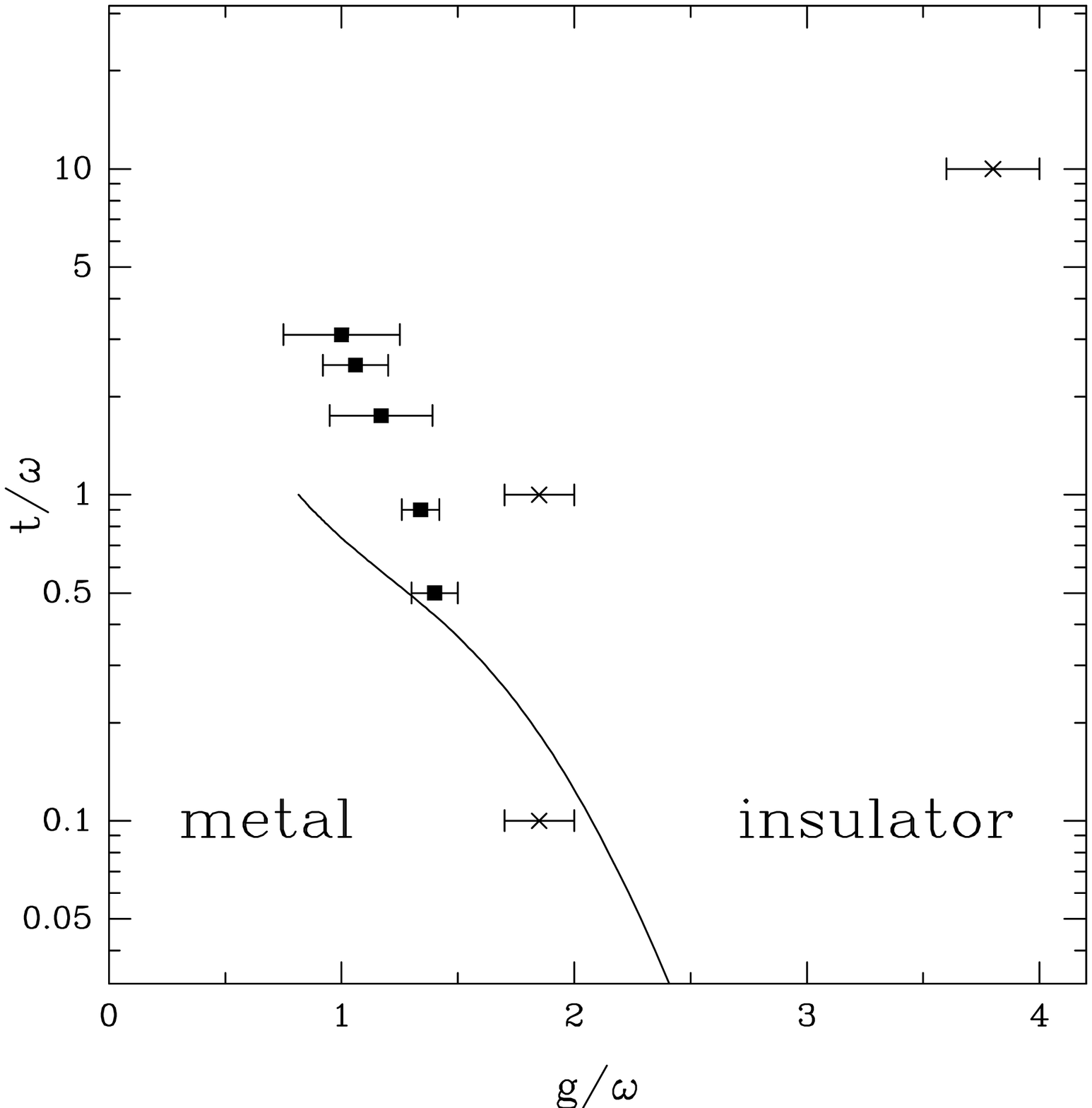}}
\begin{figure}
\caption{
Phase diagram showing the boundary between the
metallic and insulating phase.
The solid curve is the prediction of small polaron theory and the 
XXZ model
(Section \protect\ref{secpol}).
The crosses are the results of this study
and the solid squares the results of Hirsch and Fradkin 
\protect\cite{hir}.
\label{figphased}}
\end{figure}

\onecolumn
\widetext
\begin{table}
\squeezetable
\caption{
Monte Carlo results for different quantities
for $t=\omega$ and $g=1.5\omega$ and for various system sizes.
The energies are in units of $\omega$ and 
displacements in units of $(M \omega)^{-1/2}$.
 }
\begin{tabular}{cllllll}
 $N$ & $E_0(N)/N $ & $E_{-1}(N) - E_0(N) $ & $<q_i(2n_i-1)>$
&     $<q_i^2>$ & $<n_in_{i+M/2}>$ & $<n_in_{i-1+M/2}>$    \\
\tableline
   2 & -1.143 $\pm$ 0.001 &  1.158 $\pm$ 0.004 &  0.313 $\pm$
 0.005 &  0.748 $\pm$ 0.008  &  0.000
 $\pm$ 0.000 &  0.500 $\pm$ 0.000  \\
   4 & -0.895 $\pm$ 0.001  &   0.416 $\pm$ 0.004  &   0.446
$\pm$ 0.001 & 0.864 $\pm$ 0.004   &  0.260 $\pm$
0.001  &   0.120 $\pm$ 0.0004 \\
   6 & -0.868 $\pm$ 0.001 &    0.268 $\pm$ 0.009 &  
  0.470 $\pm$ 0.005 & 0.883 $\pm$  0.007 &   0.217 $\pm$ 0.001  & 
  0.244 $\pm$ 0.001 \\
   8 & -0.861 $\pm$ 0.002  &   0.200 $\pm$  0.015  & 
  0.484 $\pm$  0.002 &   0.902 $\pm$ 0.003 &   0.240
$\pm$ 0.002  &   0.229 $\pm$ 0.001 \\
  16 & -0.854 $\pm$ 0.001  &   0.064 $\pm$ 0.027  &  
 0.488 $\pm$ 0.004 &   0.904 $\pm$ 0.005 &   0.247 $\pm$ 0.002  & 
  0.246 $\pm$ 0.002 \\
\end{tabular}
\label{table1}
\end{table}

\begin{table}
\squeezetable
\caption{
Comparison of Monte Carlo results with known results for
free fermions. The ground state energy per site of
the infinite system, $\epsilon_\infty$, and 
the velocity of charge excitations, $ u_\rho$,
are normalised to their free fermion values.
The correlation exponent $K_\rho$ is
one for free fermions.
 }
\begin{tabular}{cccc}
$\displaystyle {t \over \omega  }$ &
$\displaystyle {\pi \epsilon_\infty \over 2 t }$ &
$\displaystyle {u_\rho \over 2 t }$ &
$K_\rho$ \\*[0.05in]
\tableline
0.1 & 0.999 $\pm$ 0.001 & 0.98 $\pm$ 0.03 & 1.00 $\pm$ 0.06 \\
1 & 1.000 $\pm$ 0.001 & 0.96 $\pm$ 0.02 & 1.00 $\pm$ 0.04\\
10 & 0.999 $\pm$ 0.003 & 1.00  $\pm$ 0.01 & 0.95 $\pm$ 0.02\\
\end{tabular}
\label{table2}
\end{table}


\begin{references}

\bibitem[*]{email}electronic address: ross@newt.phys.unsw.edu.au

\bibitem{gru} G. Gr\"uner, {\it Density Waves in Solids},
(Addison-Wesley, Redwood City, 1994).

\bibitem{pei} R. Peierls, {\it Quantum Theory of Solids},
(Oxford, 1955), p. 108.

\bibitem{hols} T. Holstein, Ann. Phys. {\bf 8},
325, 343  (1959).

\bibitem{hee}A. J. Heeger, S. Kivelson, J.R. Schrieffer, and W.P. Su,
Rev. Mod. Phys. {\bf 60}, 781 (1988).

\bibitem{mck}R. H. McKenzie and J. W. Wilkins,
Phys. Rev. Lett. {\bf 69}, 1085 (1992),
and references therein.

\bibitem{kim} K. Kim, R. H. McKenzie, and J. W. Wilkins,
Phys. Rev. Lett. {\bf 71}, 4015 (1993).

\bibitem{lon} F. H. Long, S. P. Love, B. I. Swanson, and R. H. McKenzie,
Phys. Rev. Lett. {\bf 71}, 762 (1993);
L. Degiorgi, St. Thieme, B. Alavi,
G. Gr\"uner, R.H. McKenzie, K. Kim, and F. Levy,
Phys. Rev. B {\bf 52}, 6203 (1995),
and references therein.

\bibitem{fra}W. P. Su, Solid State Commum.
{\bf 42}, 497  (1982); E. Fradkin and J. E. Hirsch,
 Phys. Rev. B {\bf 27}, 1680 (1983);
D. Schmeltzer, R. Zeyher, and W. Hanke,
 Phys. Rev. B {\bf 33}, 5141 (1986);
A. Auerbach and S. Kivelson,
 Phys. Rev. B {\bf 33}, 8171 (1986);
Z. B. Su, Y. X. Wang, and L. Yu, Commun. Theor. Phys.
(Beijing) {\bf 6}, 313 (1986);
J. Yu, H. Matsuoka, and W. P. Su,
 Phys. Rev. B {\bf 37}, 10367 (1988);
G. C. Psaltakis and N. Papanicolaou,
Solid State Commum.  {\bf 66}, 87 (1992);
A. Takahashi,
Phys. Rev. B {\bf 46}, 11550 (1992);
H. Zheng, Phys. Rev. B {\bf 50}, 6717 (1994);
C. Q. Wu, Q. F. Huang, and X. Sun, Phys. Rev. B {\bf
52}, 7802 (1995).

\bibitem{hir}
J. E. Hirsch and E. Fradkin, Phys. Rev. B {\bf 27}, 4302 (1983).
To compare our results with theirs set $K=M\omega^2$
and $\lambda=g(2M\omega)^{1/2}$.

\bibitem{bou} C. Bourbonnais and L. G. Caron, J. Phys. France
{\bf 50}, 2751 (1989).

\bibitem{zhe} H. Zheng, D. Feinberg, and M. Avignon,
 Phys. Rev. B {\bf 39}, 9405 (1988).

\bibitem{wu} C. Q. Wu, Q. F. Huang, and X. Sun,
Phys. Rev. B {\bf 52}, 15683 (1995).

\bibitem{voit2}
J. Voit and H. J. Schulz,                    
Phys. Rev. {\bf 36} 968, (1987).

\bibitem{fre}
J. R. Freericks, M. Jarrell, and D. J. Scalapino,
Phys. Rev. B {\bf 48}, 6302 (1993), and
references therein.

\bibitem{gub}
P. Niyaz, J. E. Gubernatis, R. T. Scalettar, and C. Y. Fong,
Phys. Rev. B {\bf 48}, 16011 (1993), and
references therein.

\bibitem{mar} F. Marsiglio, Physica C               
  {\bf 244}, 21 (1995).

\bibitem{ale}
A. S. Alexandrov, V. V. Kabanov, and D. K. Ray,
Phys. Rev. B {\bf 49}, 9915 (1994), and
references therein.

\bibitem{voit}
J. Voit, Rep. Prog. Phys. {\bf 58}, 977 (1995).

\bibitem{ric}
M. J. Rice and E. J. Mele,                       
Phys. Rev. B {\bf 25}, 1339 (1982).      

\bibitem{sp} M. Hase, I. Terasaki, and K. Uchinokura,
Phys. Rev. Lett. {\bf 70}, 3651 (1993),
and references therein.

\bibitem{frad} E. Fradkin, {\it Field Theories
of Condensed Matter Systems} (Addison Wesley, Redwood
City, 1991).

\bibitem{ben} G. Benfatto, G. Gallavotti, and J. L. Lebowitz,
Helv.  Phys.   Acta {\bf 68}, 312 (1995).

\bibitem{sho} H. B. Shore and L. M. Sander,          
 Phys. Rev. B {\bf 7}, 4537 (1973).

\bibitem{ger} B. Gerlach and H. L\"owen,
Rev. Mod. Phys. {\bf 63} 63 (1991).

\bibitem{hald}
F. D. M. Haldane,
Phys. Rev. Lett. {\bf 45}, 1358 (1980).

\bibitem{voit3}
J. Voit, J. Phys. Cond. Matter
{\bf 5}, 8305 (1993).

\bibitem{aff}
I. Affleck,
Phys. Rev. Lett. {\bf 56}, 746 (1986);
H. W. J. Bl\"ote, J. L. Cardy, and
M. P. Nightingale, Phys. Rev. Lett. {\bf 56}, 742 (1986).

\bibitem{cardy} J. L. Cardy,
J. Phys. A {\bf 17}, L385 (1984).

\bibitem{mah} G. D. Mahan, {\it Many-Particle Physics},
Second edition,
(Plenum, New York, 1990) p.285ff.

\bibitem{beni} Similar calculations were made 
by G. Beni, P. Pincus, and J. Kanamori
[Phys. Rev. B {\bf 10}, 1896 (1974)]
for the case of electrons with spin.

\bibitem{yang}
C. N. Yang and C. P. Yang,
Phys. Rev. {\bf 150}, 321, 327 (1966).

\bibitem{ham}
C. J. Hamer, J. Phys. A {\bf 19}, 3335 (1986).

\bibitem{sha}
R. Shankar, Int. J. Mod. Phys. B
{\bf 4}, 2371 (1990).

\bibitem{gia}
T. Giamarchi,
Phys. Rev. B {\bf 44}, 2905 (1991).


\bibitem{baxter}
R. J. Baxter, {\it Exactly Solved Models in Statistical
Mechanics} (Academic, London, 1982).

\bibitem{pri}
P. F. Price, C. J. Hamer and D. O'Shaughnessy, J. Phys. A
{\bf 26}, 2855 (1993).

\bibitem{cep}
D. M. Ceperley and M. H. Kalos, in ``Monte Carlo Methods in
Statistical Mechanics,'' ed. K. Binder (Springer-Verlag,
New York, 1979), p.145.

\bibitem{kal}
M. H. Kalos, D. Levesque and L. Verlet, Phys. Rev. A
{\bf 9}, 2178 (1974).

\bibitem{chi}
 S. A. Chin, J. W. Negele, and S. E. Koonin, Ann. Phys. (N.Y.)
{\bf 157}, 140 (1984).

\bibitem{hey}
  D. W. Heys and D. R. Stump, Phys. Rev. D {\bf 28}, 2067 (1983).

\bibitem{ham2}
 C. J. Hamer, K. C. Wang, and P. F. Price, Phys. Rev. D {\bf 50}, 
4693 (1994).

\bibitem{ham3}
C. J. Hamer, M. Sheppeard, Z. Weihong,
and D. Sch\"utte, submitted to
Phys. Rev. D .


\bibitem{run}
 K. J. Runge,
Phys. Rev. B {\bf 45}, 7229 (1992).

\bibitem{son}
M. Sonnek, T. Frank, and M. Wagner,
Phys. Rev. B {\bf 49}, 15637 (1994).

\end{references}
\end{document}